\documentclass[manuscript,screen,review]{acmart}

\usepackage{pifont}
\usepackage{tabularray}
\usepackage{cleveref}
\usepackage{listings}
\usepackage{subcaption}
\usepackage{tikz}
\newcommand*\circled[1]{\tikz[baseline=(char.base)]{
          \node[shape=circle,fill,inner sep=0.1pt] (char) {\textcolor{white}{#1}};}}
\newcommand{\inlinecode}{\texttt}

\setcopyright{acmcopyright}
\copyrightyear{2023}
\acmYear{2023}
\acmDOI{XXXXXXX.XXXXXXX}

\begin{document}

\title[\footnotesize{Phoenix - A Novel Technique for Performance-Aware Orchestration of Thread and Page Table Placement in NUMA Systems}]{Phoenix - A Novel Technique for Performance-Aware Orchestration of Thread and Page Table Placement in NUMA Systems}

\author{Mohammad Siavashi}
\affiliation{%
    \institution{KTH Royal Institute of Technology}
    \country{Sweden}}
\email{siavashi@kth.se}
\authornote{Work done while at Iran University of Science and Technology}
\orcid{0000-0002-2600-9025}

\author{Alireza Sanaee}
\affiliation{%
    \institution{University of Cambridge}
    \country{England}}
\authornote{Work done while at Queen Mary University of London}
\email{a.sanaee@qmul.ac.uk}
\orcid{0000-0001-6461-1650}

\author{Mohsen Sharifi}
\affiliation{%
    \institution{Iran University of Science and Technology} 
    \country{Iran}}
\email{msharifi@iust.ac.ir}
\orcid{0000-0003-4992-2500}

\author{Gianni Antichi}
\affiliation{%
    \institution{Politecnico di Milano}
    \country{Italy}}
\email{gianni.antichi@polimi.it}
\orcid{0000-0002-6063-4975}


\renewcommand{\thefootnote}{\fnsymbol{footnote}}

\renewcommand{\shortauthors}{Mohammad, et al.}

\begin{abstract}
  The emergence of symmetric multi-processing (SMP) systems with non-uniform memory access (NUMA) has prompted extensive research on process and data placement to mitigate the performance impact of NUMA on applications. However, existing solutions often overlook the coordination between the CPU scheduler and memory manager, leading to inefficient thread and page table placement. Moreover, replication techniques employed to improve locality suffer from redundant replicas, scalability barriers, and performance degradation due to memory bandwidth and inter-socket interference. In this paper, we present Phoenix, a novel integrated CPU scheduler and memory manager with on-demand page table replication mechanism. Phoenix integrates the CPU scheduler and memory management subsystems, allowing for coordinated thread and page table placement. By differentiating between data and page table pages, Phoenix enables direct migration or replication of page tables based on application behavior. Additionally, Phoenix employs memory bandwidth management mechanism to maintain Quality of Service (QoS) while mitigating coherency maintenance overhead. We implemented Phoenix as a loadable kernel module for Linux, ensuring compatibility with legacy applications and ease of deployment. Our evaluation on real hardware demonstrates that Phoenix reduces CPU cycles by 2.09x and page-walk cycles by 1.58x compared to state-of-the-art solutions.
\end{abstract}





\maketitle

\section{Introduction}

Emerge of symmetric multi-processing (SMP) systems with non-uniform memory access (NUMA) has led to years of research on both process and data placement to mitigate the NUMA effect on applications performance \cite{asymmetry-matters,page-management-design-space,balm,mitosis,vmitosis,radiant,nros,nuKSM,skip-don't-walk,is-data-migration-evil-on-nvm,inter-node-load-balancing-for-numa-systems,locality-vs-balance,nimble,numa-aware-data-placement-in-hybrid-memories,numa-aware-core-allocation,traffic-management,challenges-of-modern-numa-scheduling,trident-tiered-storage,task-based-runtime-systems,vtmm,memory-disaggregation, hydra}. Looking forward to the trending technologies, a growing number of processor sockets, and new memory technologies with diverse capabilities and capacities like NVMM are leading to more  NUMA nodes \cite{dancing-in-the-dark,nimble,vtmm}.

The NUMA effect refers to the performance degradation experienced by applications due to non-uniform memory access latencies in NUMA systems. Cross-NUMA node (remote) memory accesses can incur up to 1.4x the latency of local memory accesses, significantly impacting the performance of applications with high TLB (Translation Lookaside Buffer) miss rates and large memory footprints~\cite{mitosis}. To address this issue, several state-of-the-art solutions have proposed page table replication techniques, wherein page tables are replicated on all NUMA nodes, regardless of the application's behavior~\cite{mitosis,vmitosis,radiant,nros}. While these replication approaches increase memory locality, they introduce new challenges:

\begin{itemize}
    \item It leads to redundant replicas, limiting scalability due to the strict coherency requirements imposed by maintaining consistency across all replicas. This strict coherency necessitates holding the page table lock during updates, resulting in increased lock contention and longer page table update times. For instance, our experiments have shown a significant 24\% runtime overhead on a 4-socket machine caused by migrating a 4kb page in Mitosis~~\cite{mitosis}, a state-of-the-art solution.
    \item Accessing page table replicas can become significantly slower when memory bandwidth or cross-socket interconnections are heavily occupied. This overhead manifests as increased CPU cycles and degraded application performance. We observed up to $2.2$x runtime overhead in Mitosis compared to Linux, even on a 2-socket machine. See \S\ref{section:evaluation} for more details.
\end{itemize}

In data centers, where maximizing resource utilization is desirable and wasting CPU cycles translates to operational costs, it is imperative to minimize CPU cycles spent on coherency maintenance for redundant page table replications. Furthermore, this inefficiency significantly impacts application performance. To address this issue, this paper focuses on two key questions that aim to target the root cause:  \textit{i}) can we avoid replication by a proper thread placement in the first place? \textit{ii}) can we mitigate the number of replicas and control interference on them through low overhead mechanisms? The main focus for both questions is a solution that is compatible with legacy applications, can be easily deployed in production environments, and offers configurability.

This paper emphasizes the significance of performance, CPU cycles and page walk cycles, caused by page table and process misplacement. We contend that the integrated coordination between the CPU scheduler (thread placement) and memory manager (page table placement) plays a critical role to save CPU cycles. Our argument is as follows:

\circled{1} Current operating systems (OS) treat page table pages and data pages equally, employing the same allocation policies for both \cite{mitosis,traffic-management,challenges-of-modern-numa-scheduling,radiant}. As a result, the page allocation policy may inadvertently position the page table on a distant node compared to the process.

\circled{2} CPU scheduler decides process migration among NUMA nodes overlooking the page table location\cite{mitosis}. Once a process is migrated to a new node, AutoNUMA tries to move its data to the new node to increase localiy. This data migration updates all page table replicas. Process migration is a frequent operation in NUMA-aware schedulers. For example, VMware ESXi migrates workloads among NUMA nodes at the frequency of 2 seconds \cite{esxi-scheduler}. We measured up to $24\%$ runtime ovearhead in migrating a single 4KB data page Mitosis compared to Linux with 4 replicas.

\circled{3} Real-world workloads often experience interference in both intra-socket memory bandwidth and inter-socket connections, such as Intel UPI/QPI \cite{caladan,emba,micro-architecture-interference}. In this work, we demonstrate that the distribution of threads across all sockets by the Linux load-balancer (i.e., scheduling domains) results in interference being distributed across sockets. Consequently, page table replication leads to degraded performance due to slow page walks and replica updates caused by distributed interference (see Figure~\ref{fig:btree-motivation}).

\circled{4} Strictly coherent replication is not always necessary as many workloads experience negligible TLB miss rate, even with large memories. The OS must monitor the application's behavior to enable selective replication.

We propose Phoenix, a novel CPU scheduler and page table manager approach that mitigates NUMA effect on page-walks. Phoenix targets large-memory applications experiencing high TLB miss rates while maintaining QoS for low TLB miss applications. Phoenix employs the following techniques to maintain QoS:

\begin{itemize}
    \item \textbf{On-demand replication:} A kernel space mechanism with negligible overhead that dynamically detects processes that benefit from replication while leaving the rest to be managed by Linux completely fair scheduler (CFS).
    \item \textbf{Page table placement:} A new policy is to differentiate page table pages from data pages. This policy enables direct migration or replication of the page table based on runtime requirements.
    \item \textbf{Threads placement:} A policy to ensure that threads belonging to the same application are co-located on the same node or nodes with the fastest interconnection link. Combined with the page table placement policy, it eliminates unnecessary replicas and isolates noisy applications, minimizing interference.
    \item \textbf{Memory bandwidth management:} A hardware-assisted mechanism using Intel RDT memory bandwidth allocation (MBA) that enables Phoenix to speedup page-walk of the high-priority application by dynamically partitioning the memory bandwidth usage among high-priority and antagonistic applications running on the same node. 
\end{itemize}

Phoenix, implemented as a Linux loadable kernel module (LKM) with minimal kernel modification. It is evaluated on real hardware, demonstrating its effectiveness in improving system performance. Mitosis serves as a comparable baseline due to its similar system model. Compared to Mitosis, Phoenix achieves an average reduction of $2.09x$ in CPU cycles and $1.58x$ in page-walk cycles across various real-world workloads. Table \ref{tab:related-works-comparison} presents a comparison with state-of-the-art solutions.

We present the main contributions of this paper as follows:

\begin{itemize}
\itemsep0em
\item Based on experiments, we argue that neither thread placement nor page table replication alone can effectively mitigate the overhead of remote page-walks. However, the novelty lies in demonstrating that collaboration between both is required to maintain QoS in real-world scenarios.
\item The integration of thread placement and page table manager is proposed as a novel approach. This recognizes the need for integrating sub-systems, highlighting a new research direction in OS design.
\item The evaluation of interference management on replication solutions in data center scenarios with saturated interconnections and memory bandwidth provides valuable insights into the impact of interference on replicating OS data structures. This investigation contributes practical strategies for effective replica management.
\item The design and implementation of Phoenix on Linux, a widely used operating system in data centers, and the evaluation on real hardware with real-world applications enhance the practical relevance and applicability of the research findings.
\end{itemize}

The rest of the paper is organized as follows: \S \ref{section:motivation} covers the motivation of this research. In \S \ref{section:background}, we provide the necessary background. We present our design concepts in \S \ref{section:design} and implementation details in \S \ref{section:implementation}. Next, we evaluate the performance of Phoenix in \S \ref{section:evaluation}. Finally, we briefly discuss related works in \S \ref{section:related-works} and conclude the paper in \S \ref{section:conclusion}.

\section{Motivation} \label{section:motivation}
Improper CPU scheduling and behavior-agnostic replication employed in current solutions can result in significant performance overhead. In this section, we analyze these concerns and their impact on applications performance. We also provide the configuration details of our testbed (Table \ref{tab:system-config}) for reference.

\begin{table*}
    \centering
    \scriptsize
    \caption{Comparison of state-of-the-art solutions. vMitosis is the successor of Mitosis for virtualized environments. }
    \label{tab:related-works-comparison}
    \begin{tabular}{|l|c|c|c|c|c|}
        \hline
         System&Replication&Migration&On-demand Rep.&Contention Control&NUMA Scalability\\
         \hline\hline
         Linux&\color{red}\ding{55}&\color{red}\ding{55}&\color{red}\ding{55}&\color{red}\ding{55}&\color{red}\ding{55} \\
         \hline
         Mitosis~\cite{mitosis} / vMitosis~\cite{vmitosis}&\color{green}\ding{51}&via replication&\color{red}\ding{55}&\color{red}\ding{55}&\color{red}\ding{55} \\ 
         \hline
         NrOS~\cite{nros}&\color{green}\ding{51}&\color{red}\ding{55}&\color{red}\ding{55}&\color{red}\ding{55} & \color{green}\ding{51} \\ 
         \hline
         Phoenix&\color{green}\ding{51}&direct&\color{green}\ding{51}&\color{green}\ding{51}&\color{green}\ding{51}\\ 
         \hline
    \end{tabular}
\end{table*}

\subsection{Demand for a New CPU Scheduler}
\label{subsec:demand-for-a-new-cpu-scheduler}

The Linux CPU scheduler, inherited by Mitosis, distributes threads across all NUMA nodes to balance scheduling domains. However, this approach leads to the dispersion of threads from antagonistic applications across all NUMA nodes. Additionally, replication solutions like Mitosis allocate page tables on each NUMA node to facilitate local page-walks. This strategy results in interference on memory bandwidths and cross-socket connections, slowing down access to both local and remote replicas. Consequently, there is an increase in local page-walk cycles and the cycles required to update replicas. Our experiments on a 2-socket machine revealed up to $2.2x$ (CPU cycles) overhead in Mitosis compared to Linux.

\begin{figure*}[b]
    \centering
    \begin{subfigure}[b]{0.329\textwidth}
        \includegraphics[width=\textwidth]{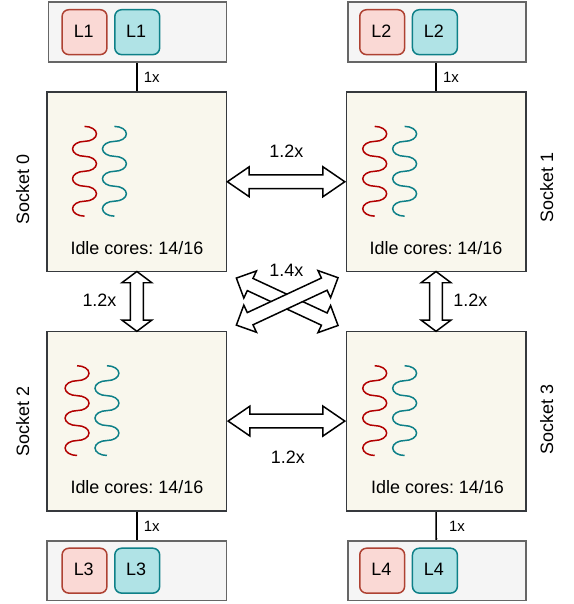}
        \caption{Linux}
        \label{subfig:linux-scheduling}
    \end{subfigure}
    \begin{subfigure}[b]{0.329\textwidth}
        \includegraphics[width=\textwidth]{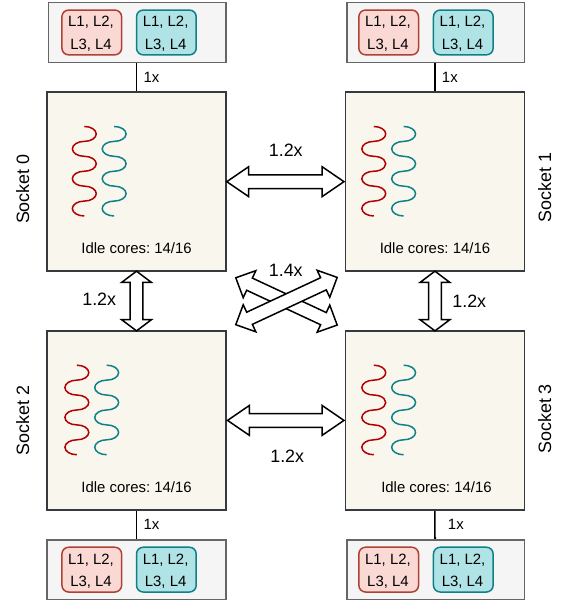}
        \caption{Mitosis}
        \label{subfig:mitosis-scheduling}
    \end{subfigure}
    \begin{subfigure}[b]{0.329\textwidth}
        \includegraphics[width=\textwidth]{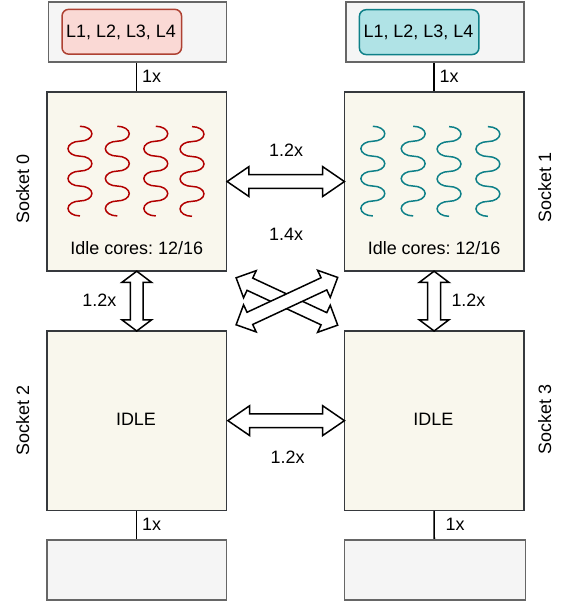}
        \caption{Phoenix}
        \label{subfig:phoenix-scheduling}
    \end{subfigure}
    \caption{Comparing thread and page table placement in Linux, Mitosis, and Phoenix running two applications. Phoenix can consolidate threads of an application on more than one node, prioritizing the nodes with the fastest interconnection for faster communication. In this figure, the basic scenario is depicted for better understanding.}
    \label{fig:scheduling}
\end{figure*}

To illustrate the problem further, Figure \ref{fig:scheduling} provides an example of the scheduling and page table placement strategies employed by Linux, Mitosis, and Phoenix when running two different applications. In this scenario, one of the applications is a low-priority process, such as a garbage collector, that saturates memory bandwidth and/or inter-socket connections. In Mitosis (Figure~\ref{subfig:mitosis-scheduling}), where each NUMA node hosts a replica, the interference hampers both local and remote accesses to replicas, significantly degrading the performance of high-priority applications. Such scenarios are prevalent in data centers, where multiple applications are packed onto a single server to maximize utilization \cite{shenango,caladan,lego-os}. To address the scheduling problem, Phoenix introduces a consolidation scheduling policy (Figure~\ref{subfig:phoenix-scheduling}) as a potential solution. Further details on the Phoenix approach can be found in \S\ref{subsec:threads-consolidation}.

\begin{figure}[!h]
    \centering
    \includegraphics[scale=0.25]{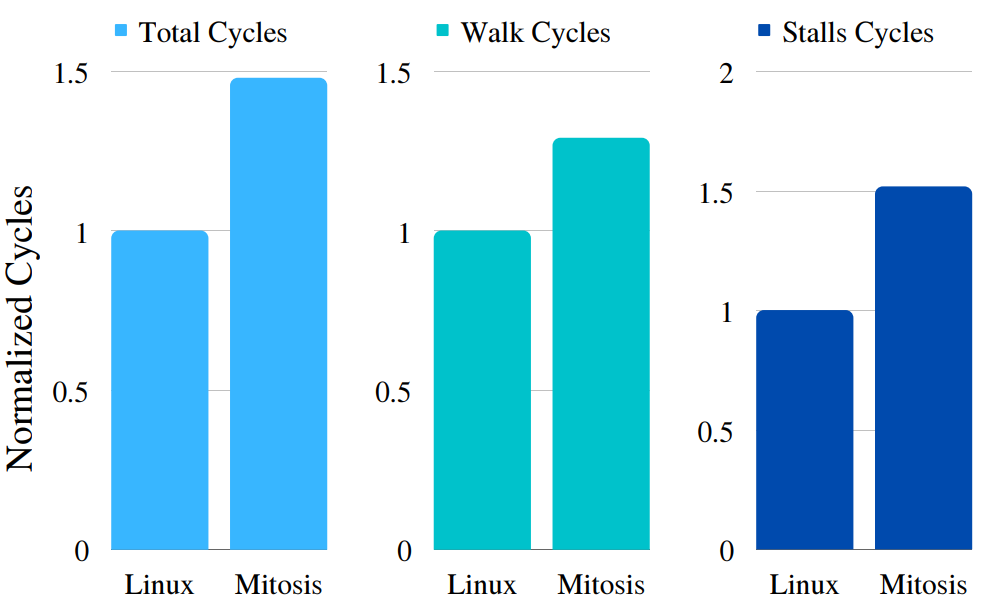}
    \caption{Performance comparison between Linux and Mitosis with 2-way replication running BTree under heavy memory bandwidth interference.}
    \label{fig:btree-motivation}
\end{figure}

We conducted an analysis of the BTree runtime parameters in both Linux and Mitosis on our 2-socket machine with 4 KB pages. To evaluate the system under a demanding scenario, we executed 32 threads of the BTree benchmark alongside 32 threads of the memory-intensive STREAM workload~\cite{stream}, which effectively saturates the memory bandwidth up to 90\%. In Figure \ref{fig:btree-motivation}, we observed a significant increase of up to $47\%$ in total CPU cycles when the memory-intensive workload saturated the memory bandwidth. To gain further insights, we employed the \texttt{perf} tool and measured the \texttt{cycle\_activity.stalls\_mem\_any} metric. Our analysis revealed that in Mitosis, the stall cycles increased by $51\%$ compared to Linux. This increase can be attributed to the replication process and the additional memory accesses related to internal implementation. Memory stalls occur when the CPU must wait for requested data, resulting in idle cycles. These idle cycles contribute to the overall increase in CPU cycles. Therefore, the rise in total cycles and page-walk cycles depicted in the figure can be attributed to stalls arising from updating the replicas.

\subsection{Behavior-Agnostic Replication}
\label{subsec:behavior-agnostic-replication}
Linux and existing page table replication solutions \cite{nros,vmitosis,mitosis} represent two opposite ends of the spectrum when it comes to page table placement. Linux, for instance, employs a single page table per process, whereas replication solutions maintain a replica per NUMA node. However, this replication approach affects the performance of certain workloads, particularly applications that frequently modify the page table. The reason behind this performance degradation lies in the \emph{behavior-agnostic replication} approach. In this paper, we contend that a more effective solution is required—one that is dynamic, takes into account application behavior, and replicates page tables on-demand—to bridge this gap.

\begin{figure}
    \centering
    \includegraphics[scale=0.25]{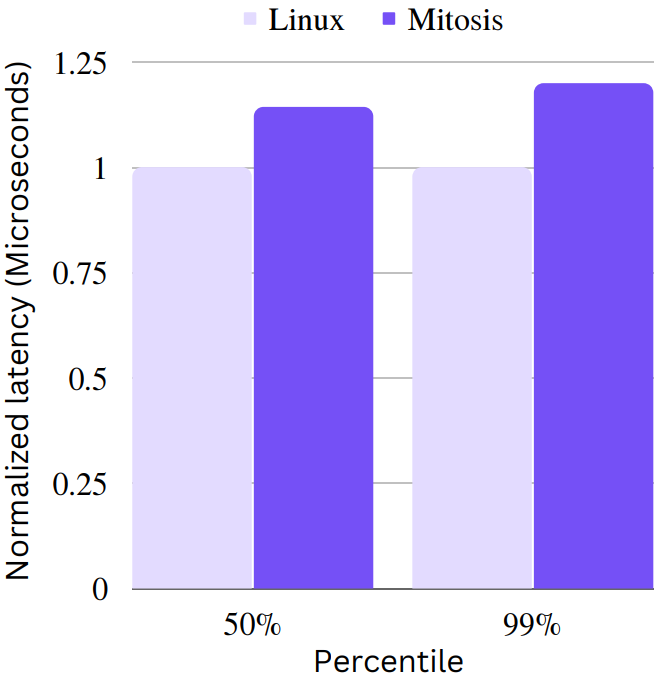}
    \caption{Apache web-server experiences higher latency with a 2-way replication. The $99\%$ percentile tail latency increased by $19.9\%$.}
    \label{fig:apache-overhead}
\end{figure}

Some applications, such as web servers (e.g., Apache \cite{apache}) and data analytic engines (e.g., MapReduce \cite{mapreduce,evaluating-mapreduce}), frequently perform virtual memory operations that modify the page table \cite{didi,optimizing-tlb-shootdown-algorithm,latr}. To investigate this behavior, we conducted an experiment where 16 threads of Apache served a default 12 KB static page under a load of $5000$ concurrent requests, totaling $50000$ requests. The results, depicted in Figure \ref{fig:apache-overhead}, demonstrate that the Mitosis replication solution leads to a $19.9\%$ increase in the $99^{th}$ percentile tail latency on a 2-socket machine.

\begin{figure*}[b]
     \centering
     \begin{subfigure}[b]{0.33\textwidth}
         \centering
         \includegraphics[width=\textwidth]{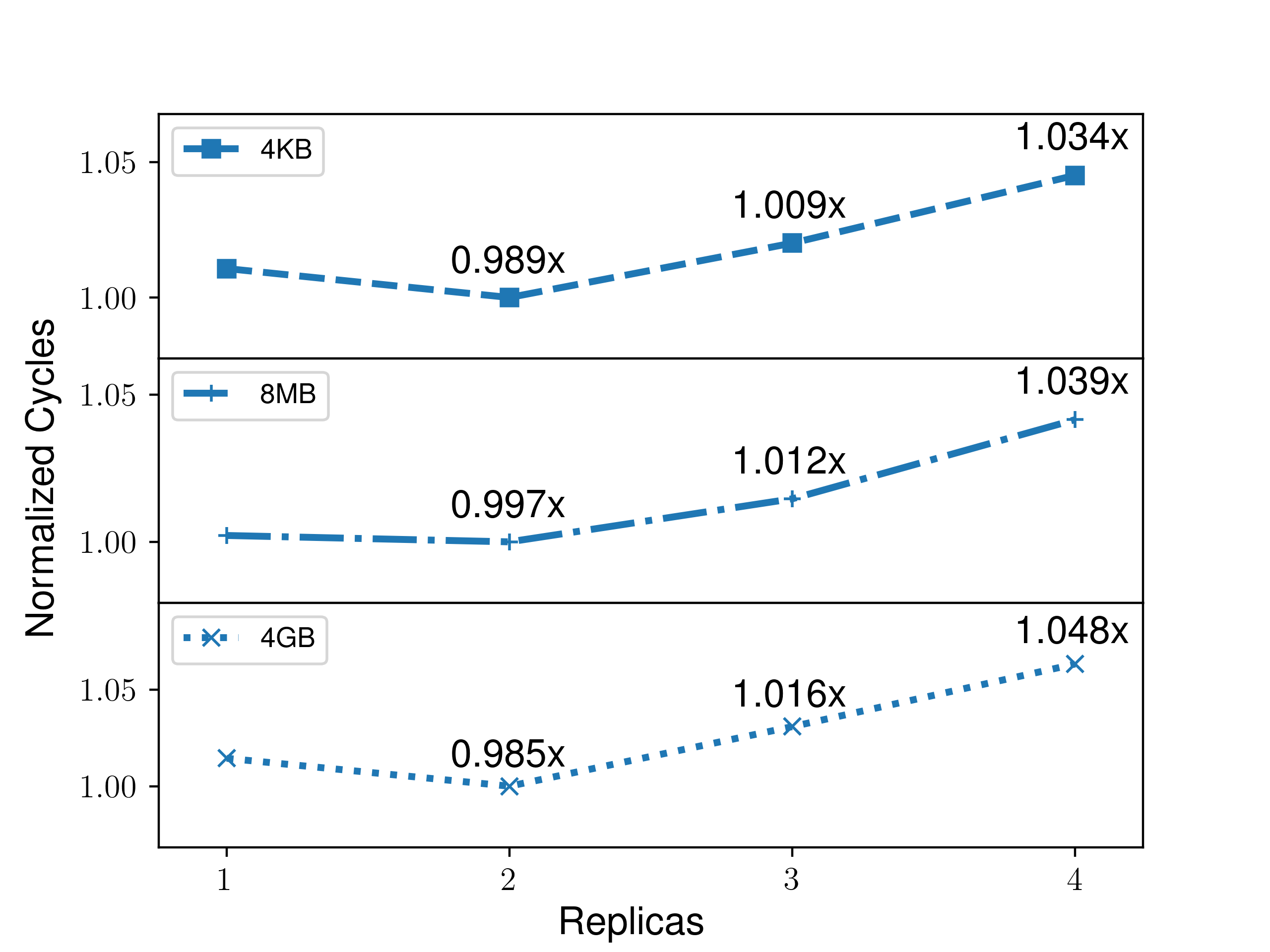}
         \caption{mmap}
         \label{fig:mmap-overhead}
     \end{subfigure}
     \hfill
     \begin{subfigure}[b]{0.33\textwidth}
         \centering
         \includegraphics[width=\textwidth]{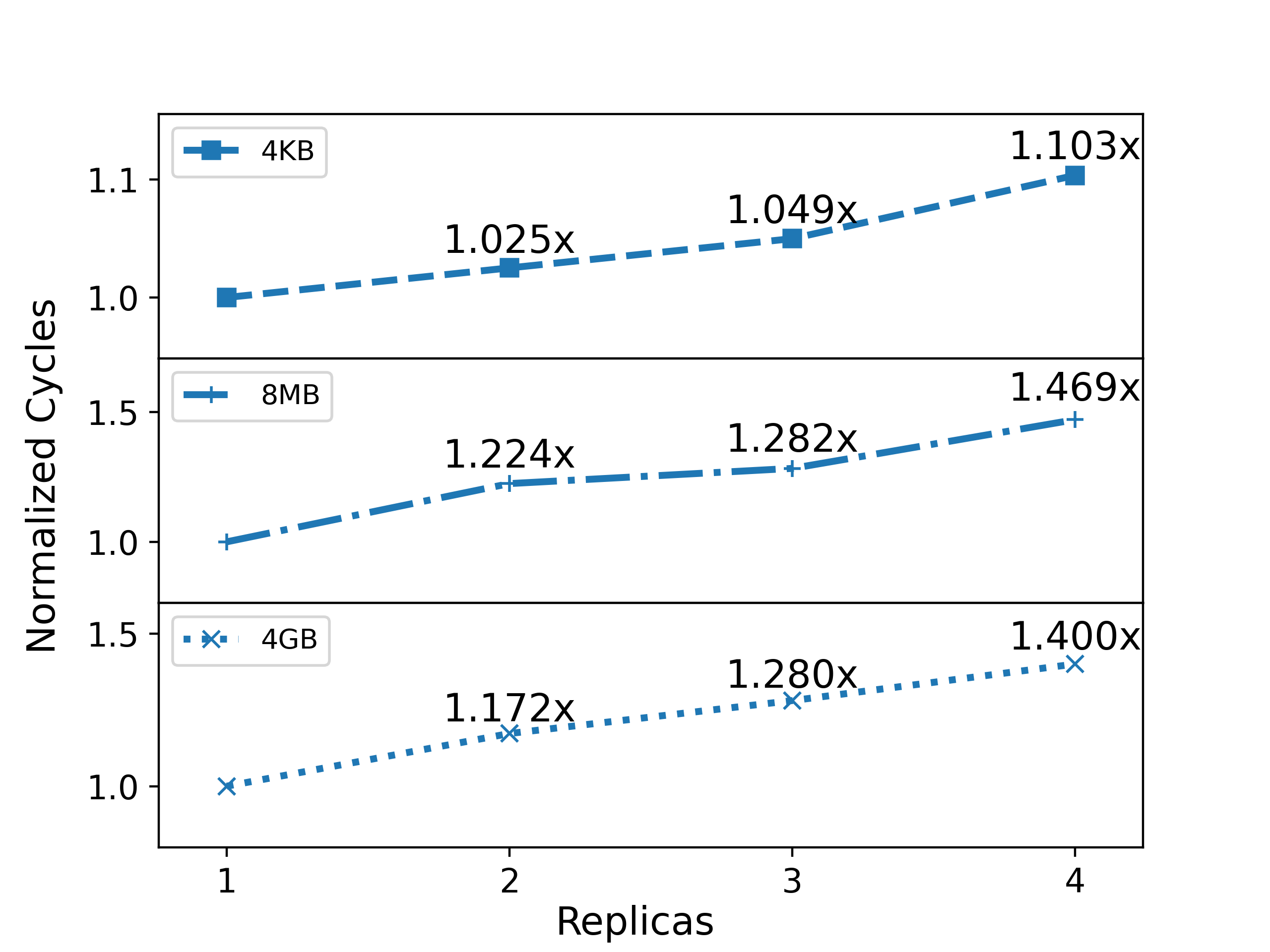}
         \caption{munmap}
         \label{fig:munmap-overhead}
     \end{subfigure}
     \hfill
     \begin{subfigure}[b]{0.33\textwidth}
         \centering
         \includegraphics[width=\textwidth]{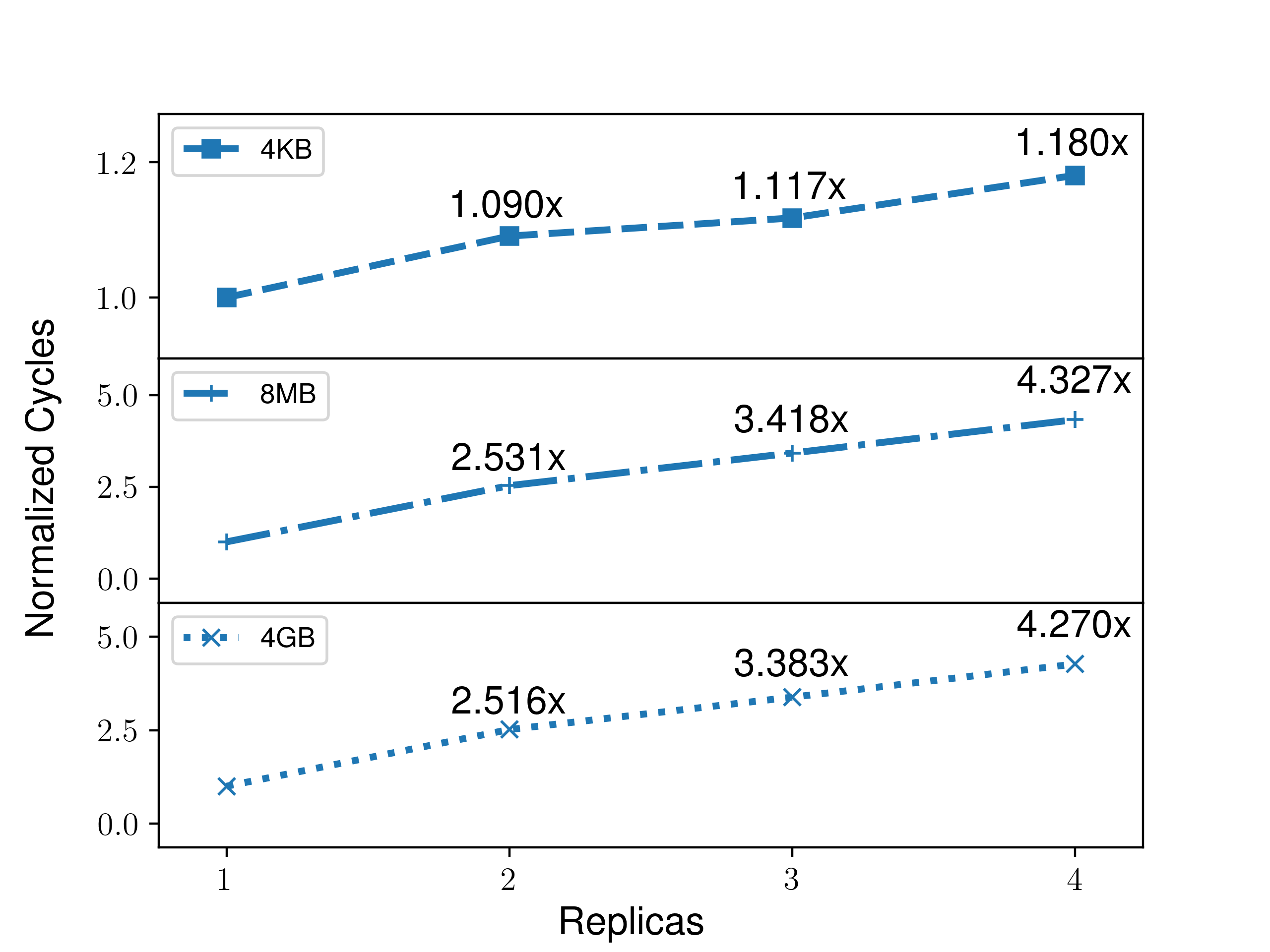}
         \caption{mprotect}
         \label{fig:mprotect-overhead}
     \end{subfigure}
        \caption{Overhead of replication on memory management system calls as the number of replicas grows. Mitosis uses page cache that causes mmap line to drop when replication turned on. }
        \label{fig:syscalls-overhead}
\end{figure*}

As an additional example, we employed the Metis benchmark suite, which comprises high-performance MapReduce benchmarks \cite{analysis-of-linux-scalability-on-many-cores,radixvm,optimizing-mapreduce}. By utilizing \texttt{strace}, we were able to identify that the \emph{Metis wrmem} (in-memory inverted index calculation) extensively utilizes \inlinecode{mremap} and \inlinecode{mprotect} when the input size exceeds 36 GB, resulting in a memory footprint of 190 GB. Notably, nearly all address translations hit the TLB with a d-TLB miss rate of less than $0.05\%$. Table~\ref{tab:wrmem-syscalls} illustrates the overhead imposed by Mitosis on system calls while profiling \emph{wrmem} on our 2-socket machine, running 16 threads of wrmem. Additionally, we developed three micro-benchmarks that specifically target the most common virtual memory operation system calls, demonstrating how overhead escalates with the number of replicas. Figure~\ref{fig:syscalls-overhead} depicts the results of our micro-benchmarks conducted on a 4-socket machine. In this experiment, we measured a runtime overhead ranging from $3.2\%$ to $7.8\%$.

\begin{table}[!h]
\caption{Overhead of replication on virtual memory syscalls in Wrmem (MapReduce)}
\label{tab:wrmem-syscalls}
\begin{tabular}{@{}lcc|cc@{}}
\toprule
         & \multicolumn{2}{c}{Linux} & \multicolumn{2}{c}{Mitosis} \\ \midrule
         & \% time    & usecs/call   & \% time     & usecs/call    \\
mremap   & 53.54      & 53           & 55.17       & 61            \\
mprotect & 40.50      & 21           & 36.44       & 23            \\
mmap     & 4.66       & 25           & 7.21        & 28            \\
munmap   & 0.90       & 30           & 0.29        & 30            \\ \bottomrule
\end{tabular}
\end{table}

\section{Background} \label{section:background}
This section covers the necessary background required for the rest of the paper.

\subsection{Page Tables}
Modern OS with virtual memory use page tables to map virtual addresses to physical addresses. Linux uses a 4-5 levels radix tree per process on x86\_64 architecture \cite{radixvm}. For each address translation query, a page-walk is initiated by a page miss handler that is aware of the page table structure. A full page-walk requires 4-5 memory accesses. It goes up to 24 in virtualized environments. As a result, processor vendors use multi-level TLBs to cache page table entries (PTEs). To further optimize, processors leverage methods such as multiple page-walk units, out-of-order execution, TLB prefetcher, special-purpose page-walk caches (PWC), and general-purpose data caches \cite{every-walk-is-a-hit}. These optimizations effectively reduce 4 memory accesses to 1.39 on average (max 2.5 in random access workload) and from 24 to 4.4 in virtualized environments \cite{compendia}. 

\begin{figure}[!h]
    \centering
    \includegraphics[scale=0.25]{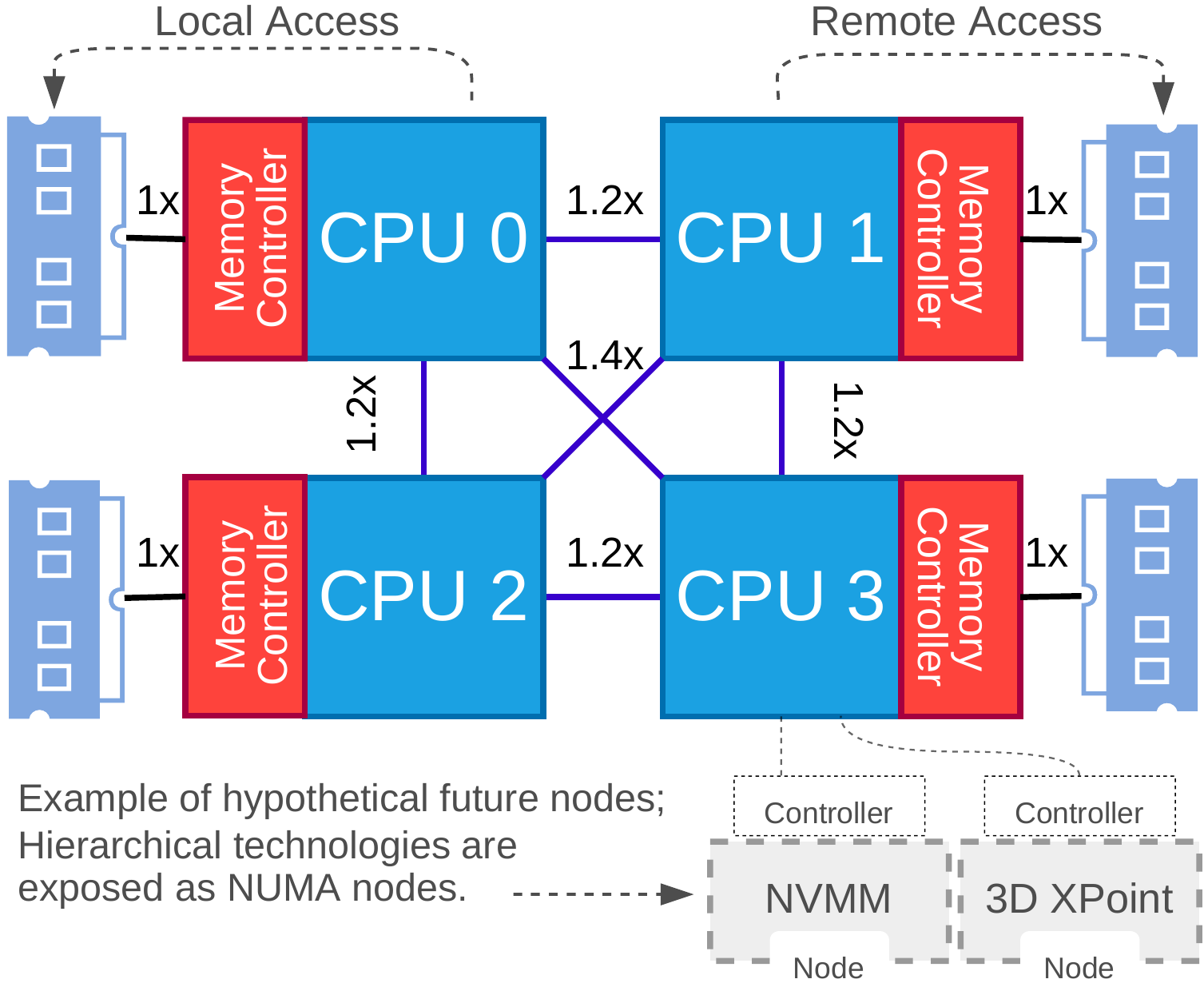}
    \caption{Example of an SMP NUMA system}
    \label{fig:NUMA-architecture}
\end{figure}

Large-memory applications allocate page tables that are beyond the coverage of TLB caches \cite{big-memory-servers}. It increases the TLB miss rate and page-walk overhead. The overhead exacerbates when the process initiating the page-walk is located on a distant node relative to the page table due to cross-socket memory access latency \cite{mitosis}. Figure \ref{fig:NUMA-architecture} shows the difference between local and remote memory accesses and their latencies.

\subsection{NUMA Architecture}
Scale-up demand in the last decade has led to the prevalence of multi-socket machines in data centers. In modern multi-socket x86\_64 architectures, each processor is connected to a memory cluster managed by a dedicated memory controller. As Figure \ref{fig:NUMA-architecture} illustrates, each memory controller unit forms a NUMA node \cite{locality-vs-balance}.  In each node, access to the local memory is the fastest, while accessing memory on a distant node takes 1.2-1.4x through cross-socket interconnections. Today, some large-memory machines utilize tiered memory technologies such as Intel Optane. Tiered and heterogeneous memory systems are becoming a popular feature of current and future machines. Different memory technologies, such as 3D XPoint, disaggregated memory, and high bandwidth memory (HBM), each form a NUMA node with different properties \cite{nimble,vtmm}. Increasing sockets and NUMA nodes have attracted many researchers regarding scheduling and memory management, indicating that new scheduling and memory management policies are necessary as the number of NUMA nodes continues to grow \cite{trident-harnessing-architectural-resources}.

\subsection{Data Placement in NUMA Systems}

\begin{figure}[b]
     \centering
     \begin{subfigure}[]{0.49\columnwidth}
         \centering
         \includegraphics[scale=0.65]{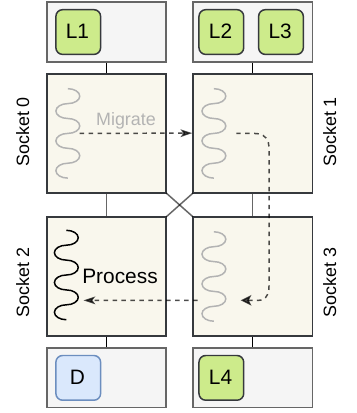}
         \caption{First-Touch}
         \label{fig:first-touch}
     \end{subfigure}
     \hfill
     \begin{subfigure}[]{0.49\columnwidth}
         \centering
         \includegraphics[scale=0.65]{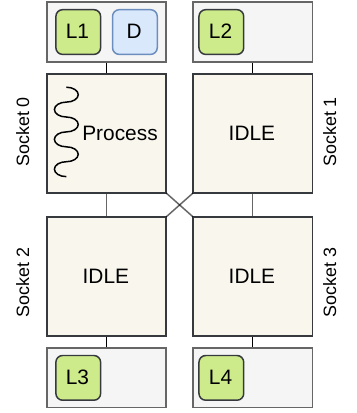}
         \caption{Interleave}
         \label{fig:interleave}
     \end{subfigure}
    \caption{Example of Linux page allocation policies. \textit{Li} denotes the level of the page table radix tree and \textit{D} denotes data.}
    \label{fig:page-allocation-policies}
\end{figure}

Modern operating systems implement NUMA-aware data placement policies. In Linux, there are two main policies for initial page allocation. The default policy, known as \emph{first-touch}, allocates pages on the NUMA node that issues the allocation request. Alternatively, the \emph{interleave} policy distributes pages evenly among NUMA nodes using a round-robin approach. Figure \ref{fig:page-allocation-policies} illustrates an example of these policies that are applied to page tables. While these policies benefit certain workloads \cite{locality-vs-balance}, studies have shown that adaptive page management policies can significantly enhance performance \cite{adaptive-page-migration}.

Linux utilizes the AutoNUMA subsystem to perform \emph{lazy} migration of data pages that are frequently accessed from a distant NUMA node. This involves periodically unmapping target pages from the process page table. When a process attempts to access an unmapped page, a page fault occurs, and the kernel intervenes to identify the process responsible for the access and decide whether to migrate the page or not. However, our micro-benchmarks reveal that Mitosis introduces runtime overheads of $16\%$ and $24\%$ for a single page migration on our Skylake and Sandy Bridge machines, respectively. While AutoNUMA effectively handles data migration, it lacks support for page table migration \cite{mitosis}. In contrast to Linux, Phoenix supports both process and page table migration in an integrated manner, mitigating the overheads associated with the lazy migration approach~\cite{mitosis}.

\subsection{NUMA-Aware Scheduling}
NUMA scheduling is a challenging problem since it requires a deep understanding of scheduling and memory management. Additionally, after a decade, the community is still exploring to shed light on new aspects of problems that a NUMA scheduler must address. In this research, we present a novel approach to address the NUMA scheduling and memory management problems by exploring page table awareness in this domain.

The Linux kernel utilizes the \texttt{struct task\_struct} data structure to store attributes and runtime information for each schedulable entity. Within the kernel, both processes and threads are referred to as tasks, without a distinction between them. Tasks can create additional tasks, forming task groups. On NUMA systems, the Linux scheduler aims to balance scheduling domains by distributing tasks of a task group across different nodes \cite{inter-node-load-balancing-for-numa-systems}.

\begin{figure}[!h]
    \centering
    \includegraphics[scale=0.85]{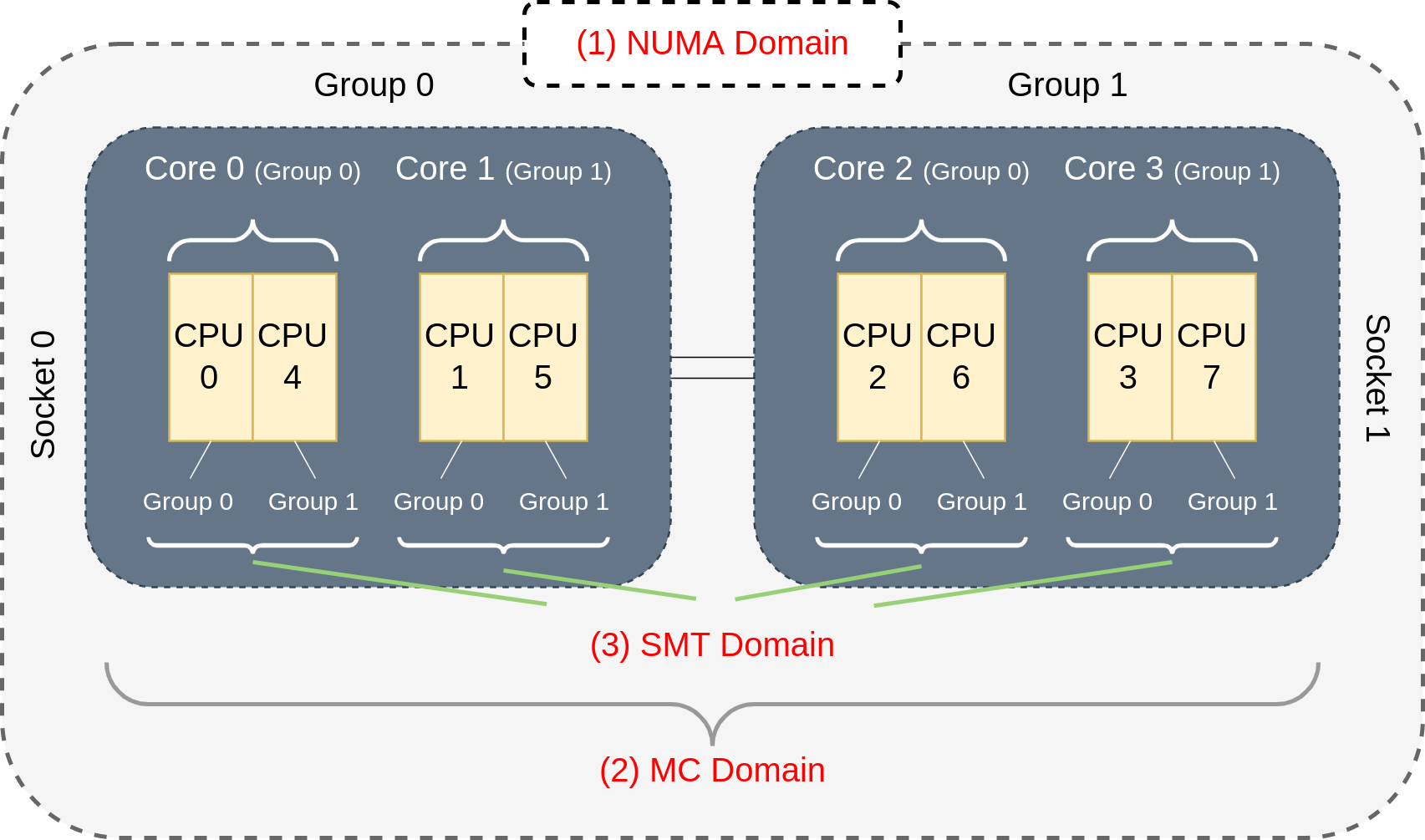}
    \caption{Linux scheduling domains and scheduling groups for a 2-socket machine. Balancing is done in the following order: NUMA domain, MC (Memory Controller) Domain, and SMT (Simultaneous Multithreading) Domain.}
    \label{fig:scheduling-domains}
\end{figure}

To optimize scheduling decisions and minimize contention for shared resources, Linux organizes processors into logical hierarchies known as \emph{scheduling domains}. These domains are formed based on the shared physical resources, such as the Last-Level Cache (LLC) \cite{decade-of-wasted-cores}. As Figure \ref{fig:scheduling-domains} illustrates, in a 2-socket machine, three scheduling domains are created, with the top-level domain being the NUMA domain. The NUMA domain groups cores within the same NUMA node, the SMT domain groups cores that share a physical core but have different hardware threads, and the MC domain groups cores that share the same memory controller. This hierarchical organization facilitates load balancing, which starts from the top-level NUMA domain, ensuring a relatively equal distribution of threads across NUMA nodes.

CFS tolerates $25\%$ load imbalance between sockets, meaning a socket can contain 12 threads while the other contains 15 \cite{battle-of-schedulers}. However, the load-balancer tries to achieve a balance in initial threads placement. In \S \ref{section:motivation}, we argue that this thread placement policy enforces redundant page table replication while also exacerbating page-walk overhead in the presence of interference.

\section{Phoenix Design} \label{section:design}
\begin{figure}[!t]
    \centering
    \includegraphics[scale=0.6]{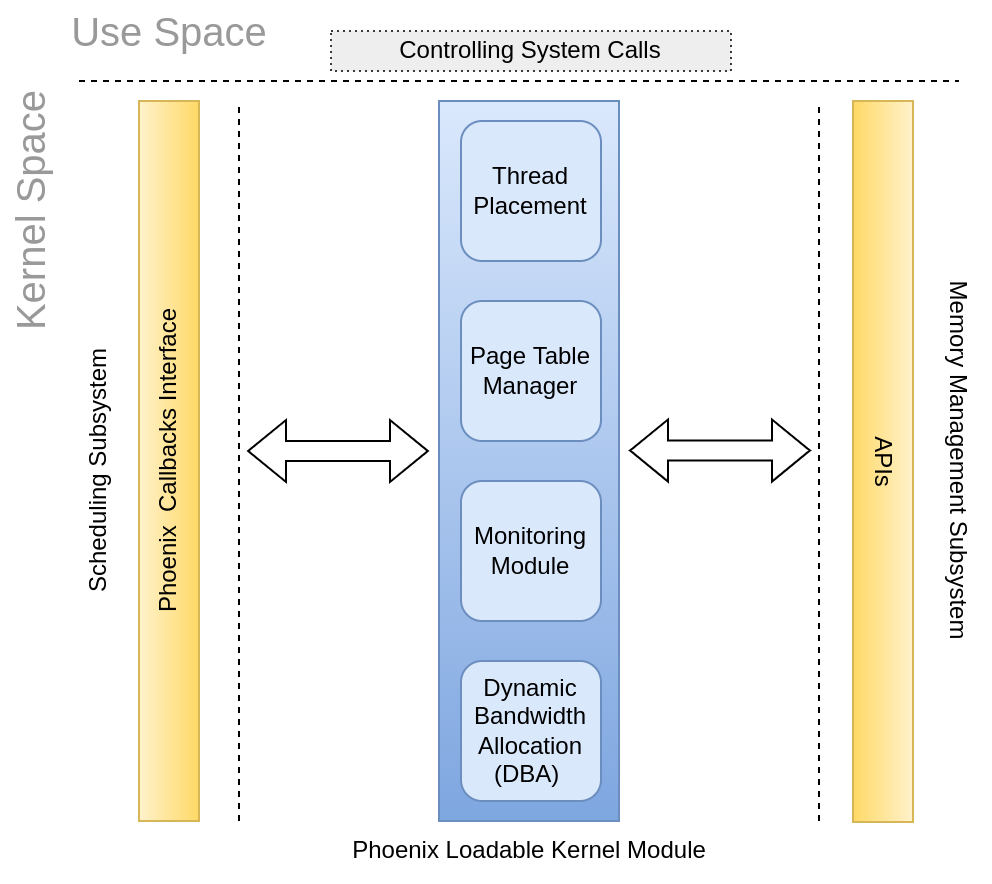}
    \caption{Phoenix abstract architecture}
    \label{fig:phoenix-design}
\end{figure}

Phoenix primary design concept is to retain threads and their page table together in an integrated fashion. The purpose is to reduce the overhead of remote page-walks while mitigating the unnecessary overheads of eager replication in real-world data center scenarios. There are three main challenges in Phoenix design: \emph{i)} a low overhead mechanism to monitor process behavior for on-demand replication and interference detection, \emph{ii)} a policy that serves as an integrated coordinator between the scheduler and memory manager, and \emph{iii)} a mechanism to enable transparent and fast page table replication and direct migration. Figure \ref{fig:phoenix-design} illustrates an abstraction of Phoenix architecture. In this section, we detail the design of Phoenix. 

\subsection{Challenges and Design Decisions}
\noindent \textbf{Target Workloads:} In order to maximize utilization, data centers often run a mix of different types of applications on a single server. For instance, Microsoft Bing utilizes this approach on $90,000$ of its servers, and Google compute clusters typically run an average of 8 applications per machine~\cite{shenango}. This co-location of diverse workloads leads to increased contention for shared resources. Phoenix is designed to address these scenarios where multiple applications are packed on a single server, with each application utilizing only a fraction of the total available resources.

\noindent \textbf{Flexibility and Transparency:} Phoenix offers flexibility and transparency by seamlessly detecting and managing its target workloads at runtime, while allowing Linux to handle the remaining workloads. The core logic of Phoenix is implemented as a Loadable Kernel Module (LKM) using established kernel APIs. This approach ensures straightforward maintenance and deployment procedures without requiring system downtime. Furthermore, Phoenix provides essential configuration options through the \inlinecode{procfs} interface, enabling easy customization of the system. A key advantage of Phoenix is its compatibility with legacy applications, as it does not require any modifications at the application level.

\noindent \textbf{No Slowdown, No Conflict: } Thread placement and updating a task's statistics on context-switches are on the critical path \cite{decade-of-wasted-cores}. Coordination between the scheduler and the memory manager should not slow down the critical path. Meanwhile, a decision must be made in an integrated fashion to minimize overhead. Besides, it should not conflict with existing Linux policies to avoid undefined behaviors. We ensure no conflict occurs at the critical path by reusing high-level kernel APIs. Additionally, we maintain per CPU and task statistical variables to trace the system's state for decision makings.

\subsection{On-Demand Replication} 
Mitosis extends \inlinecode{numactl} to manually enable replication for desired applications. Lack of on-demand replication at the runtime makes replication solutions inefficient and impractical in environments where applications can execute or terminate anytime \cite{radiant}. Certain applications do not derive any benefits from replication as their page tables are already well-cached. On-demand replication, however, possesses the capability to enhance performance without causing any slowdown for such applications. 

Applications with high page-walk overhead replicate whenever required, enhancing their overall performance. Furthermore, high page-walk overhead is not always due to application behavior itself. We observed that contention on inter-socket interconnections, memory bandwidth, and TLBs could dramatically increase page-walk overhead. Hence, solely monitoring PMCs is usually misleading. Instead, we resolve interference before replicating to ensure that replication is the last thing we do.

\subsection{Threads Consolidation} 
\label{subsec:threads-consolidation}
In scenarios where multiple applications are co-located on a single server, each application consumes a portion of the overall system resources. Traditionally, balancing threads between NUMA nodes was considered a reasonable design choice in the absence of dynamic resource partitioning technologies. However, the Phoenix system takes a different approach by attempting to group threads from the same process onto a single node, referred to as the "home node."

If the threads or data of an application exceed the available cores or memory of the home node, the home node is expanded to include additional nodes connected via the fastest interconnection. This placement policy in Phoenix yields performance improvements over Linux and Mitosis for several reasons: \circled{1} it eliminates redundant page table replicas, \circled{2} it reduces the need for migrating workloads across NUMA nodes, \circled{3} it mitigates the overhead of cross-socket TLB shootdown operations \cite{ptlbmalloc2}, and \circled{4} it enhances cache hotness.

In scenarios where applications can be placed on disjoint sets of NUMA nodes, consolidation isolates the applications. However, a question arises when multiple applications share the same node: how can interference be managed? The feasibility of interference management on a single socket can be achieved with emerging resource partitioning technologies such as Intel RDT (Resource Director Technology) \cite{intel-rdt}.

\subsection{Memory Bandwidth Management} Processes are assigned to the least occupied node with respect to memory bandwidth usage and the number of available cores. A memory-heavy antagonistic application may share a \emph{home node} with a high-priority application. Since memory bandwidth throttling significantly increases page-walk and replicas updating latency, Phoenix uses Intel Memory Bandwidth Allocation (MBA \cite{intel-mba}) to throttle memory bandwidth usage of the antagonistic application dynamically.

\subsection{Hyper-Threading TLB Contention} 
To investigate TLB sharing mechanisms on the same physical core, we developed a small program consisting of two worker threads that repeatedly access the same memory region. We ensured that the program executed on specific cores using CPU pinning. We executed the program on different combinations of logical and physical cores, monitoring TLB-related performance counters using the perf tool.

When executing the program on separate physical cores, the TLB load misses remained at a similar level, suggesting that TLB sharing did not occur within the same physical core. Interestingly, executing the program on the same physical core resulted in the highest number of TLB load misses and page walks, indicating a lack of TLB sharing.

The results from our experiments suggest that TLB sharing does not occur on sibling hyper-threads even for threads of the same application. Hence, our experiments show no advantage in sharing sibling logical cores, even between two threads of the same process, regarding the TLB miss rate. 

We measured $51.4x$ higher d-TLB miss resulting in $30.5x$ higher page-walk cycles when co-locating wrmem ($0.04\%$ d-TLB miss in isolation) with our micro-benchmark ($99.9\%$ d-TLB miss). Also, memory-intensive workloads slow down page-walks of the application located on sibling logical core \cite{holmes}. Furthermore, in order to not under-utilize the machine and adhere to the targeted data center scenario, as well as ensure a fair comparison with Mitosis, hyper-threading is not disabled in our experiments, while still suggesting its disabling to avoid TLB and PWCs contentions.

\section{Implementation} \label{section:implementation}
In this section, we explain the implementation details of Phoenix for $x86\_64$ architecture in Linux. Although the problem still exists in latest Linux kernel, we implemented Phoenix on kernel v5.4 to compare with Mitosis. Phoenix's main logic is an LKM, and our minimal kernel changes can be easily patched on recent kernels.

NUMA-aware scheduling and memory management are not easy. In other words, a destructive solution requires extensive kernel changes that may introduce significant instability, making them less usable in production and usually incompatible with legacy applications. With Phoenix, we focus on practical production solutions without extensive kernel changes. We stand on the shoulders of giants by reusing existing kernel APIs from memory management and scheduler subsystems to implement Phoenix.

\subsection{Monitoring Processes Behavior}
\label{monitoring-processes-behavior}
On-demand replication requires a low-overhead passive mechanism to monitor process behavior at the kernel level. The performance monitoring unit (PMU) provides hardware performance counters that only impose the negligible overhead of setting a register. However, it is not a straightforward operation at the kernel level as it requires setting, resetting, and reloading registers at the correct positions \cite{pmctrack}. To benefit from the flexibility of LKMs, we added a set of function pointers as callbacks (Listing~\ref{lst:callbacks}) in the kernel. Our LKM implements callbacks to provide per-task monitoring with high accuracy. We measured less than $1\%$ difference in the values of the counters compared to \inlinecode{perf} tool. We also added three members to \inlinecode{task\_struct}. First, \inlinecode{struct pmc\_sample\_struct} to store performance counters values which is allocated in \inlinecode{on\_fork} callback when the task is initialized. The second is a \inlinecode{boolean} to enable/disable Phoenix, and the third is an array pointing to the page table replicas. We refer to the first index of this array as \emph{home node} and the rest as \emph{allowed nodes}.

\lstset{basicstyle=\small}
\begin{lstlisting}[language=C, linewidth=\columnwidth, breaklines=true, caption=Events callbacks in kernel, captionpos=b, label={lst:callbacks}]
struct phoenix_callbacks {
 int (*on_fork)(unsigned long, struct task_struct *);
 void (*on_cs_out)(struct task_struct *, int);
 void (*on_cs_in)(struct task_struct *, int);
 void (*on_scheduler_tick)(struct task_struct *, int);
 void (*on_exec_task)(struct task_struct *);
 void (*on_free_task)(struct task_struct *);
 void (*on_exit_task)(struct task_struct *);
 void (*on_run_rebalance_domains)(struct rq *);
};
\end{lstlisting}

Intuitively, the d-TLB miss ratio seems to be a good indicator for on-demand replication \cite{mitosis}. Our observations show that a process can have a relatively low d-TLB miss rate while still suffering from high page-walk overhead. Page-walks could be affected by conditions such as QPI/UPI or memory bandwidth interference. As a result, we monitor total CPU cycles and page-walk cycles. The threshold on which Phoenix triggers the on-demand replication could be set from user-space by \inlinecode{sysctl}. In our study, we set a fixed threshold of $10\%$ for triggering replication. This choice was made primarily for feasibility purposes and to ensure consistent replication across all experiments. It is important to note that determining an optimal and dynamic threshold requires further in-depth research, which extends beyond the scope of our current study.

Phoenix relies on a traditional technique to measure the memory bandwidth usage of task groups. It monitors \inlinecode{LLC\_MISSES} and multiplies it by the \inlinecode{CACHELINE\_SIZE} to estimate the memory bandwidth usage \cite{caladan}. The shortcoming of this approach is that it does not count the prefetch misses. To have a more accurate estimation \inlinecode{CAS\_COUNT.RD} should be involved in the equation. However, for our prototype development, the abovementioned approach is fast and easy to implement while giving enough accuracy to detect bandwidth throttling.

\subsection{Threads Consolidation}
Linux allows implementing customized policies using high-level APIs such as \inlinecode{sched\_setaffinity()}. This function is used to bind a task to a specific core. It enables fine-grain control over threads placement while abstracting away the details of low-level scheduling. To implement our prototype, we rely on this function to selectively schedule threads on a subset of processors and nodes. By using \inlinecode{sched\_setaffinity()}, Phoenix ensures that each process runs on the core that is closest to its memory location, thereby minimizing the number of remote memory accesses.

This makes Phoenix highly stable while leaving the scheduling within the selected subset of processors to CFS.

\emph{Fork} system-call starts a new process by copying the parent process address space, while \emph{exec} replaces the already allocated address space with data from another image. As a result, to start a new process, \emph{fork} and \emph{exec} are called in order, while threads are executed immediately at \emph{fork} since they only need a copy of their parent address space. We add a callback in \inlinecode{sched\_fork()} to allocate and initialize per-task data structures. Another callback is placed in \inlinecode{sched\_exec()} to do the initial task placement. This function is called early before the task is assigned to any \emph{runqueue}, making it a proper place for initial task placement. To avoid expensive \inlinecode{for\_each\_core} loop in the scheduler hot path, we use \inlinecode{DEFINE\_PER\_CPU} macro to define a per CPU \inlinecode{struct} that keeps track of total memory bandwidth usage. 

Based on this information, Phoenix initially locates processes on the node with the least memory bandwidth usage and more idle cores prioritizing memory bandwidth. We set this node and \emph{home node} for this process. When the process spawns new threads, they inherit the \emph{home node} of the parent. If placing the spawned thread on the \emph{home node} does not oversubscribe the processor, the thread is placed there. Otherwise, Phoenix selects the closest node (fastest QPI/UPI) considering memory bandwidth and idle cores.


\subsection{Page Table Placement}

The page table placement policy follows the \emph{home node} and \emph{allowed nodes}. Initially, Phoenix places the page table on the \emph{home node}. Page table allocation functions, \inlinecode{pmd\_alloc\_one()}, \inlinecode{pud\_alloc\_one()}, \inlinecode{p4d\_alloc\_one()}, \inlinecode{pte\_alloc\_one()}, and \inlinecode{\_pgd\_alloc()} are modified to initially allocate page table pages on the \emph{home node}. This avoids scattering the page table on different nodes caused by Linux policies.

Page table migration is complex as it requires fixing the page table entries to ensure that address mappings are correct. Additionally, page table migration is prone to data race, as a page under migration might be modified by a page fault. We do not consider migrations or replication bottlenecks of our system since both are infrequent operations, especially when threads consolidation is enabled. Besides, migrating a page table page takes only a few microseconds \cite{vmitosis}. Mitosis acquires
\inlinecode{spinlock page\_table\_lock} while replicating the page table. This lock blocks access to the entire data structure and increases the replication overhead. Phoenix further reduces this overhead using the fine-grain per-table lock (\inlinecode{ptl}) for migrating PTE and PMD tables. We ignore PGD migration since it is usually well-cached \cite{radiant}. Replication and migration are practically the same, with the only difference being that the previous replica must be freed in migration. Like Mitosis, we use page caches to reserve enough pages to avoid out-of-memory during migrations. Since our primary focus is not reducing the memory capacity overheads, we leave this for future work.

Hardware-assisted page-walk starts by reading the \inlinecode{CR3} register. Hence, the address of replicas must be stored. We modify \inlinecode{mm\_struct} to add an array of pointers to \inlinecode{pgd\_t} of each replica. This array is used to load the local \inlinecode{pgd\_t} address on each node to \inlinecode{CR3} register for the page-walk to be done on the local replica.

Coherency must be maintained among replicas. A naive approach is to traverse the tree structure of each replica to update the target entry on every page table update imposing a significant overhead. For this, we adapt the Mitosis approach to implementing a low-overhead solution. Like Mitosis, we implement a circular linked list between replicas pages by adding a pointer to the next replica page in \inlinecode{struct page}. When a page table page is updated on a replica, Phoenix traverses the pointer to the  \inlinecode{next} replica page and updates until all replicas are updated. We assume this operation could be further optimized by lazy or on-demand mechanisms \cite{nros}. This needs further investigation.

\subsection{Memory Bandwidth Management}
Memory bandwidth might be saturated by a low-priority antagonistic process such as garbage collectors, degrading the performance of a high-priority application located on the same node. Bandwidth interference increases page-walk cycles as well as latency to access data. In such scenarios, Phoenix uses Intel RDT to throttle the memory bandwidth of the antagonistic application. 

In our prototype implementation of Phoenix we aimed on investigating the feasibility of our approach. In order to experiment scenarios where memory bandwidth interference occurs due to low-priority antagonistic processes, such as garbage collectors, we deliberately throttled the memory bandwidth of these processes to the maximum possible level. This allowed us to observe the effects on page-walk cycles and data access latency.

However, it is important to acknowledge that our current approach of maximum bandwidth throttling is not necessarily the most dynamic or fair solution for bandwidth allocation. We recognize that achieving a more dynamic and equitable allocation of memory bandwidth requires further research and development. Therefore, we consider this aspect of bandwidth allocation as an avenue for future work, where we can explore and implement more sophisticated strategies to dynamically manage memory bandwidth, ensuring fairer resource distribution among competing processes.






\section{Experimental Evaluation} \label{section:evaluation}
In this section, we present the performance evaluation of the Phoenix system using a set of real-world applications. Our system setup is described in Table \ref{tab:system-config}, and the workloads used for the experiments are listed in Table \ref{tab:workloads}. For all experiments, AutoNUMA is enabled. 

It is worth noting that the page tables of the applications discussed in Section \ref{subsec:behavior-agnostic-replication} are well-cached in the cache hierarchy. As a result, Phoenix's on-demand replication does not perform replication for these workloads, as it would only degrade their performance. Therefore, Phoenix performs on par with Linux in these cases, and we do not repeat these experiments in this section. Instead, we focus on providing an extensive evaluation of applications with high TLB misses.

We evaluate the overhead imposed by UPI and memory-bandwidth saturation on page table replication. Moreover, we show that a proper thread placement and dedicated page allocation policy for page table pages and on-demand replication improve performance for various workloads. Since we only have 2 sockets, we first run benchmarks that can be divided into separate sockets to show upper-bound improvement. Next, we evaluate scenarios where applications overlap on a node.

    
\subsection{Experimental Setup} 
We evaluate our system on 2 servers presented in Table \ref{tab:system-config} both running Ubuntu 20.04 with kernel v5.4. We had limited access to a rather outdated 4-socket Sandy-bridge server with small memory on each node and unsupported Intel RDT, which hindered us from running most of our experiments. As a result, we ran experiments on a modern 2-socket Skylake machine. However, we use the older machine to demonstrate that the overhead trend increases almost linearly as more replicas are added to the system.
\begin{table}[!h]
    \centering
    \caption{System configuration}
    \label{tab:system-config}
    
    \begin{subtable}[t]{\columnwidth}
        \caption{Hardware}
        \centering
        \begin{tabular}{|c|c|c|}
            \hline
            Processor Model & Cores & Memory \\
            \hline
             Intel Xeon Gold 6142 x 2 & 16, 2 HT & 384 GB \\
             \hline
             Intel Xeon E5-4620 x 4 & 8, 2 HT & 128 GB \\
             \hline
        \end{tabular}
    \end{subtable}
    
    \begin{subtable}[!b]{\columnwidth}
        \caption{System Settings}
        \centering
        \begin{tabular}{|c|c|c|}
            \hline
            Linux Kernel & DVFS & HT\\
            \hline
            5.4& Performance & on\\
            \hline
        \end{tabular}
    \end{subtable}
\end{table}


Despite Phoenix support for transparent huge-pages (THP), we omit its evaluations for brevity. It is because huge-pages enable larger memory coverage by TLB caches that decrease page-walk cycles. This only makes Phoenix treat the application as an application that does not benefit from replication due to low page-walk overhead. As this is regardless of the page size, we see no benefit for literature in adding extra experiments.

In our evaluation, we made a deliberate decision not to replicate the specific scenarios used in the evaluation of Mitosis. These scenarios involve manual pinning of both the antagonistic and benchmark applications to a single NUMA node, followed by the manual migration of the targeted application to a new node while keeping the antagonistic application pinned to its initial node. While this approach enables the manual isolation of the antagonistic application from the targeted application, we believe that it does not accurately reflect real-world scenarios.

In contrast, Phoenix takes a different approach by operating within the default settings of the system, without the need for manual pinning. This approach allows Phoenix to provide a more holistic and realistic evaluation compared to the corner case evaluation employed by Mitosis. By working within the system's default settings, Phoenix captures the actual behavior and performance in typical, everyday scenarios. This ensures that the evaluation results of Phoenix are more representative of real-world use cases, enhancing the relevance and applicability of our findings.

\subsection{Baseline Performance}
\begin{figure}[!h]
    \centering
    \includegraphics[scale=0.5]{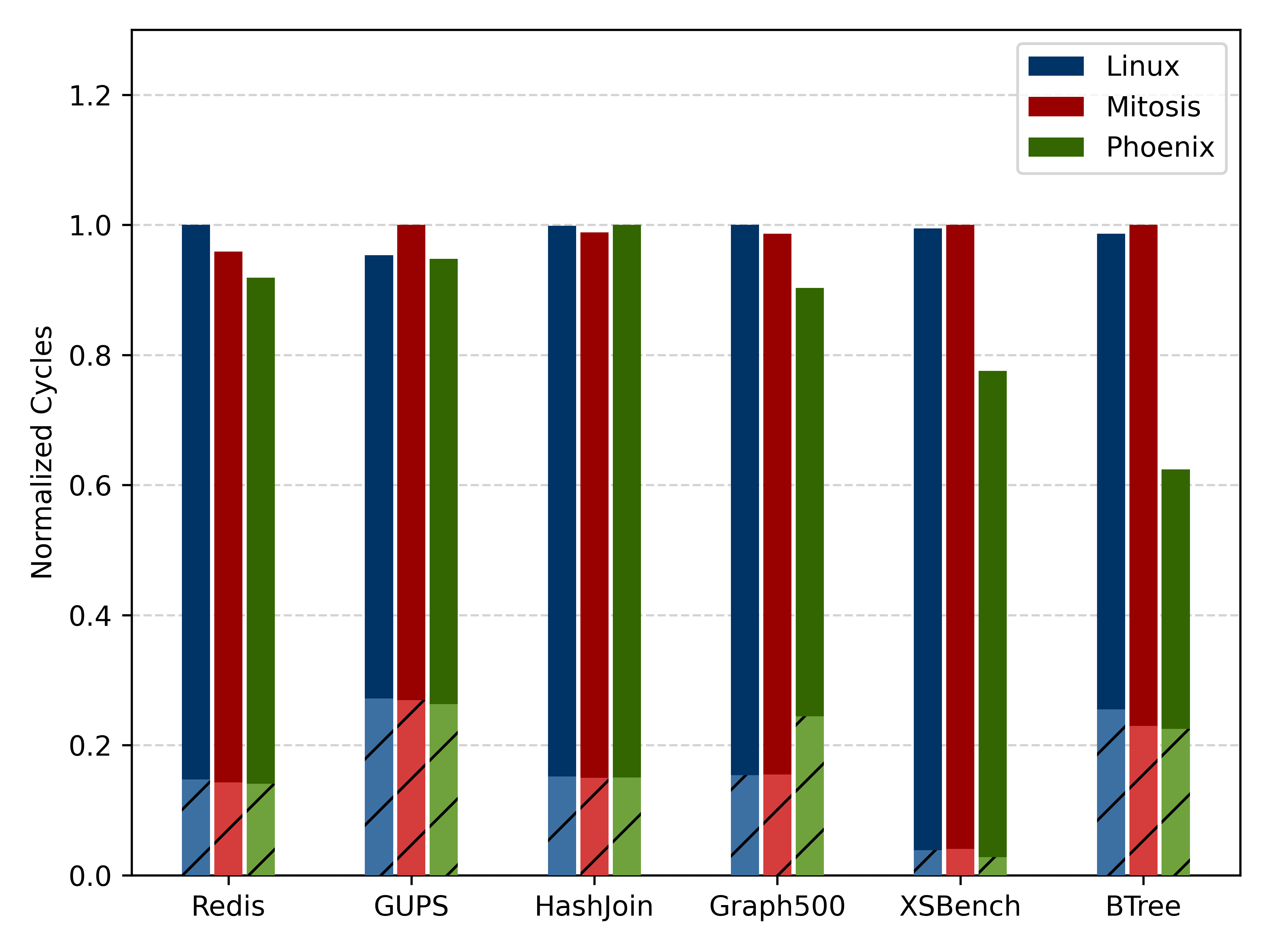}
    \caption{Performance comparison of applications running on Linux, Mitosis, and Phoenix without interference from an antagonistic application. The hashed parts represent the page-walk cycles.}
    \label{fig:no-interference}
\end{figure}

To establish a baseline and evaluate the runtime overhead of Phoenix in a normal scenario without interference from antagonistic applications, we conducted individual executions of each application on the machine using the default operating system settings. The results, depicted in Figure \ref{fig:no-interference}, reveal the performance differences compared to Linux, Mitosis, and Phoenix.

Under normal conditions, Mitosis exhibits an average overhead of $0.30\%$ in terms of total CPU cycles compared to Linux, while also reducing page-walk cycles by $1.12\%$. This can be attributed to the implementation overheads of Mitosis outweighing its benefits when severe interference is absent. On the other hand, Phoenix demonstrates a significant reduction in CPU cycles with average savings of $26.86\%$ and $27.24\%$ compared to Linux and Mitosis, respectively. Furthermore, in terms of page-walk cycles, Phoenix improves performance by $2.87\%$ compared to Linux and $1.72\%$ compared to Mitosis.

\begin{table}[]
    \centering
    \footnotesize
    \caption{List of workloads used in experiments.}
    \label{tab:workloads}
    \begin{tblr}{X[0.8] X[3.6] X[0.5]}
        \hline
        \textbf{Workload} & \textbf{Description} & \textbf{Data}\\
        \hline
        Redis & A single-thread key-value store \cite{redis}. Key size = 25, value size = 64, elements = 256M, 100\% reads. & 75 GB\\
        \hline
        GUPS & A HPC benchmark that performs integer random update \cite{gups}. & 64 GB \\
        \hline 
        HashJoin & A hash-table lookup benchmark used in large-scale applications. Elements = 128M. & 35 GB  \\
        \hline
        XSBench & A Monte Carlo neutron transport algorithm \cite{xsbench}. Threads = 32, p factor 15000000, g factor 180000. & 85~GB \\
        \hline
        BTree & An Index lookup benchmark used in large-scale applications. Threads = 32, elements = 1572M. & 150~GB \\
        \hline
        Graph500 & Generate, compress and search on large graphs \cite{graph500}. Threads = 32, scale factor = 28, edge factor = 32. & 113~GB \\
        \hline
        Wrmem & A high-performance MapReduce in-memory inverted index calculation from Metis benchmark 
        \cite{mapreduce,analysis-of-linux-scalability-on-many-cores}. Threads = 16, Input size = 36 GB.& 190~GB  \\
        \hline
        Apache & A widely used web server designed for modern OSs. We use it to serve its default 12 KB static web page. & - \\
        \hline
        STREAM & A memory-heavy benchmark were used to create the worst-case scenario of memory bandwidth interference \cite{stream}. & 23~GB\\
    \end{tblr}
\end{table}

It is important to note that Mitosis did not run its evaluations without interference, and therefore its baseline performance was not considered in its paper. These findings highlight the advantageous performance of Phoenix and its negligible overhead in a normal setting, showcasing substantial performance improvement compared to both Linux and Mitosis.

\subsection{Comparison}
\label{subsec:comparison-to-other-systems}

In a typical data-center setting, many applications co-exist in parallel in a given time. In Linux, threads of these applications are co-located on all nodes and migrate to balance the NUMA domain. In this case, the data page could be migrated to the accessing thread's locality or remain remote due to access pattern. Both cause contention on inter-socket interconnection and memory bandwidth, leading to higher latency in updating page table replicas and even local page-walks.

To create the worst-case interference scenario, we co-locate real-world applications with STREAM. Our analysis shows that the UPI interconnection is saturated if we run STREAM with interleave memory allocation policy. Otherwise, if we run it with the first-touch policy, the stress shifts from UPI to the memory bandwidth.

\begin{figure*}[!h]
    \centering
    \begin{subfigure}[t]{0.49\textwidth}
        \includegraphics[width=\columnwidth]{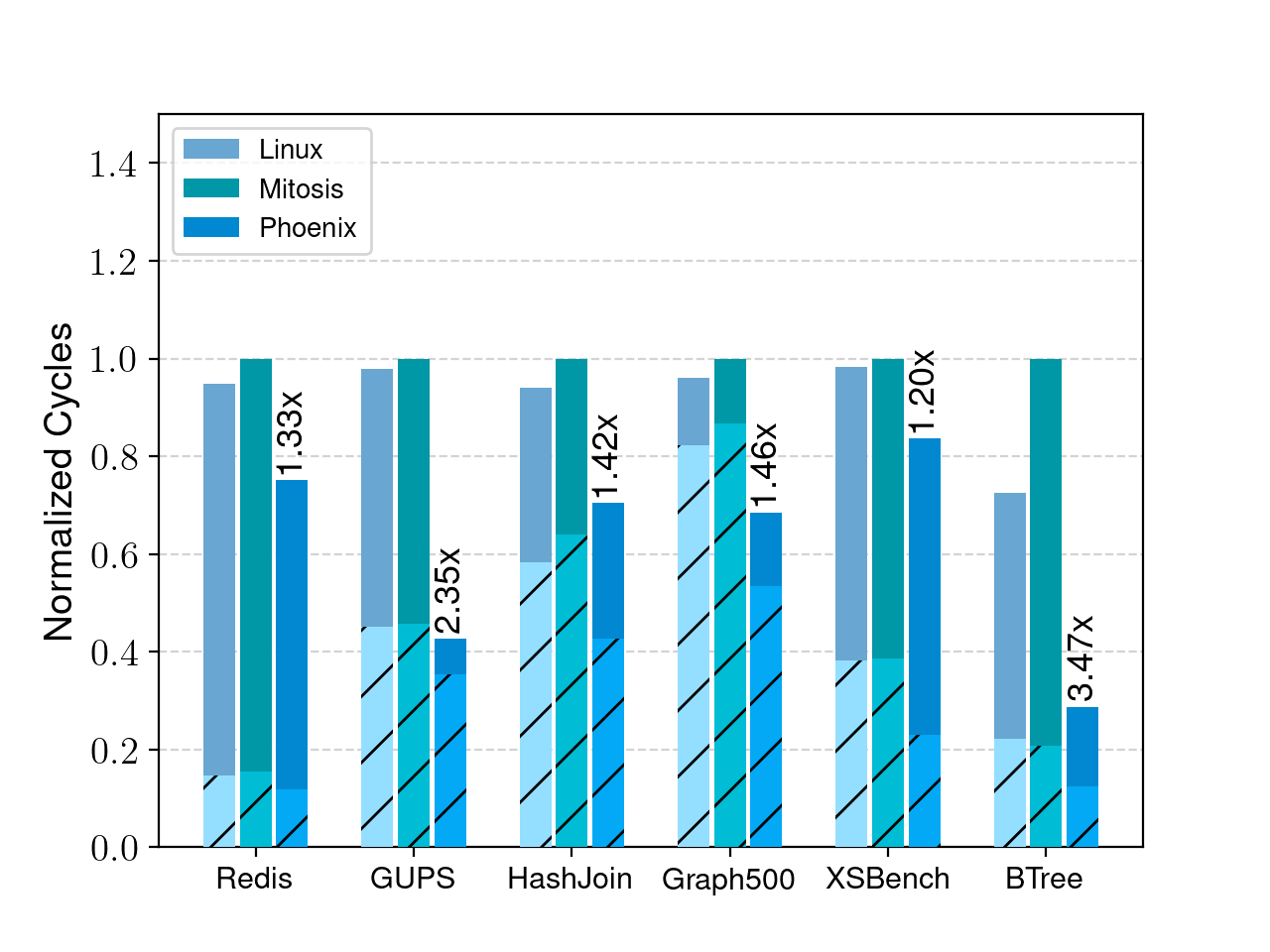}
        \caption{UPI interference}
    \end{subfigure}
    \begin{subfigure}[t]{0.49\textwidth}
        \includegraphics[width=\columnwidth]{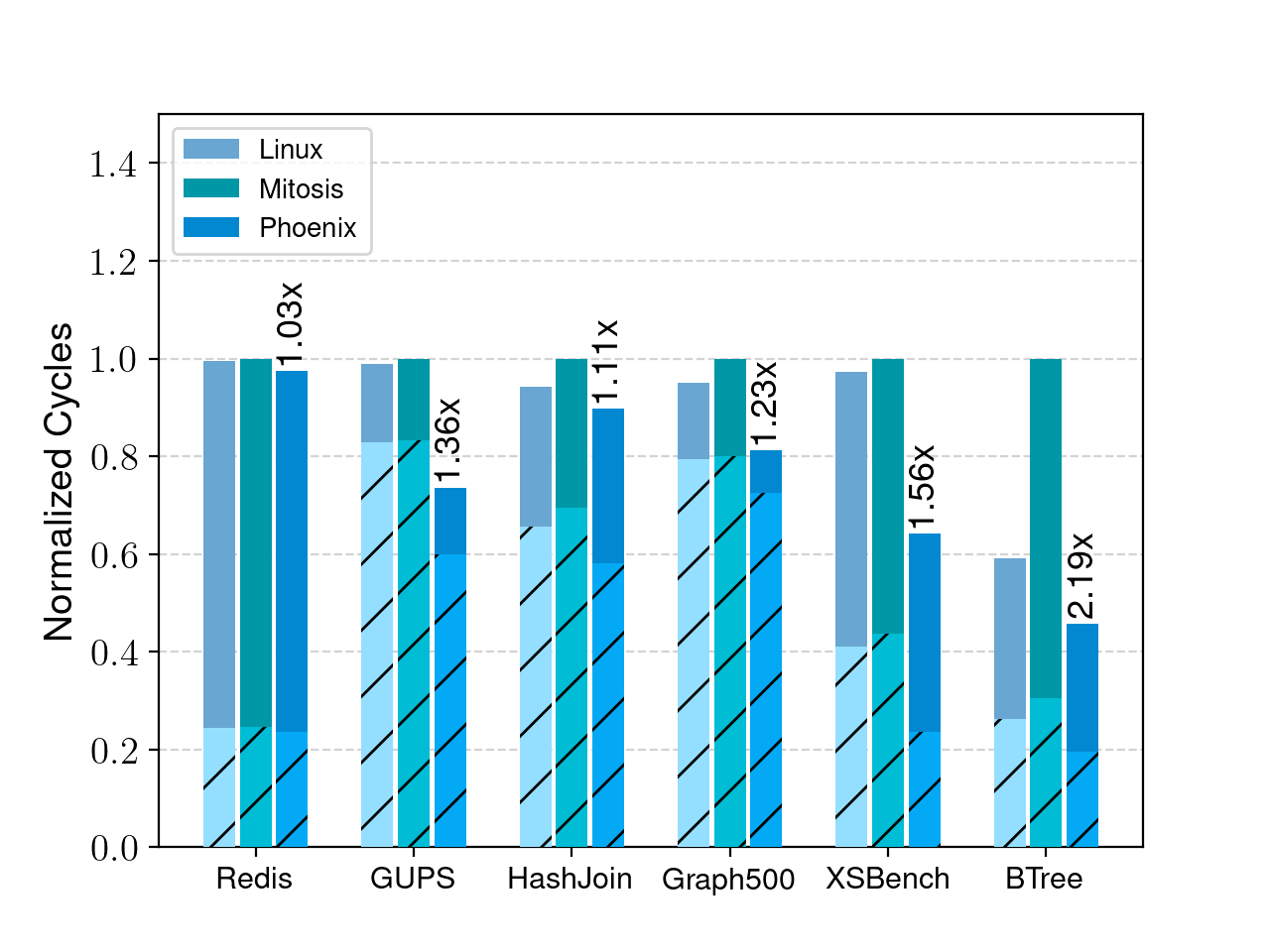}
        \caption{Memory bandwidth interference}
    \end{subfigure}
    \caption{Performance evaluation of multi-application scenario. Hashed part shows page-walk cycles. The annotated enhancement is compared to Mitosis.}
    \label{fig:interference-benchmarks}
\end{figure*}

Results of our evaluations are presented in Figure \ref{fig:interference-benchmarks}. At a glance, it can be inferred that under this typical data center scenario, Mitosis eager replication performs worst compared to Linux and Phoenix. Maintaining replicas becomes even more expensive, degrading the overall performance even on a 2-socket machine. In other words, replicas that are supposed to enhance performance become the system's performance bottleneck. Meanwhile, when UPI is throttled, Phoenix outperforms Linux and Mitosis, on average, by $1.66x$ and $1.87x$ respectively. When memory bandwidth is throttled, it enhances performance by $1.36x$ in Linux and $1.41x$ in Mitosis due to threads consolidation.


\subsection{Memory Bandwidth Management} 
\label{subsec:mba-evaluation}
A decade ago, no hardware resource partitioning existed when NUMA awareness was added to the Linux scheduler. Hence, the design decisions to balance NUMA nodes seem reasonable for generic workloads. With emerging technologies such as Intel RDT that allow dynamic resource partitioning and fine-grain resource monitoring capabilities, better design decisions can be made with negligible overhead. 

\begin{figure*}[t]
    \centering
    \begin{subfigure}[t]{0.49\textwidth}
        \includegraphics[width=\columnwidth]{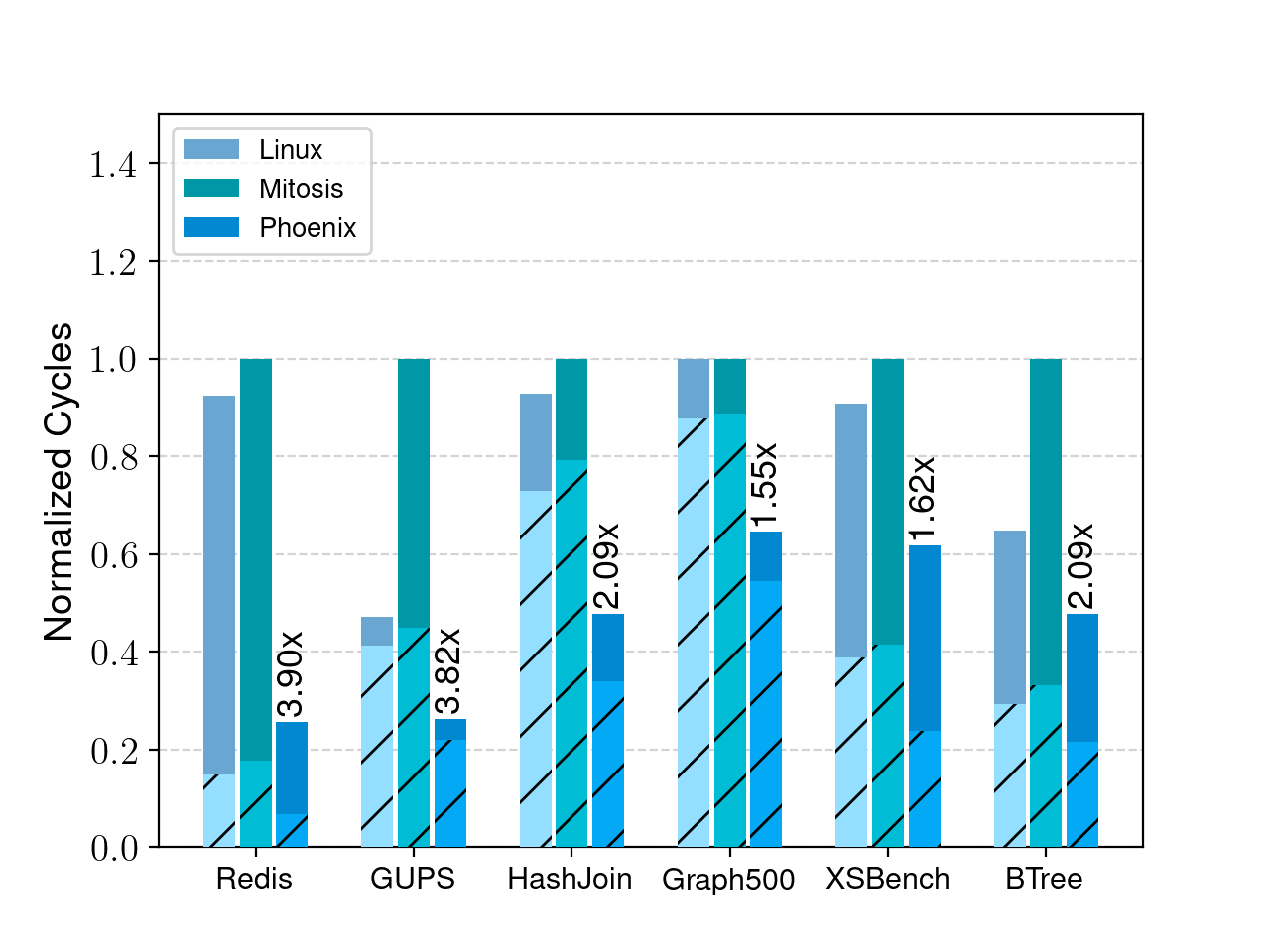}
        \caption{UPI interference}
    \end{subfigure}
    \begin{subfigure}[t]{0.49\textwidth}
        \includegraphics[width=\columnwidth]{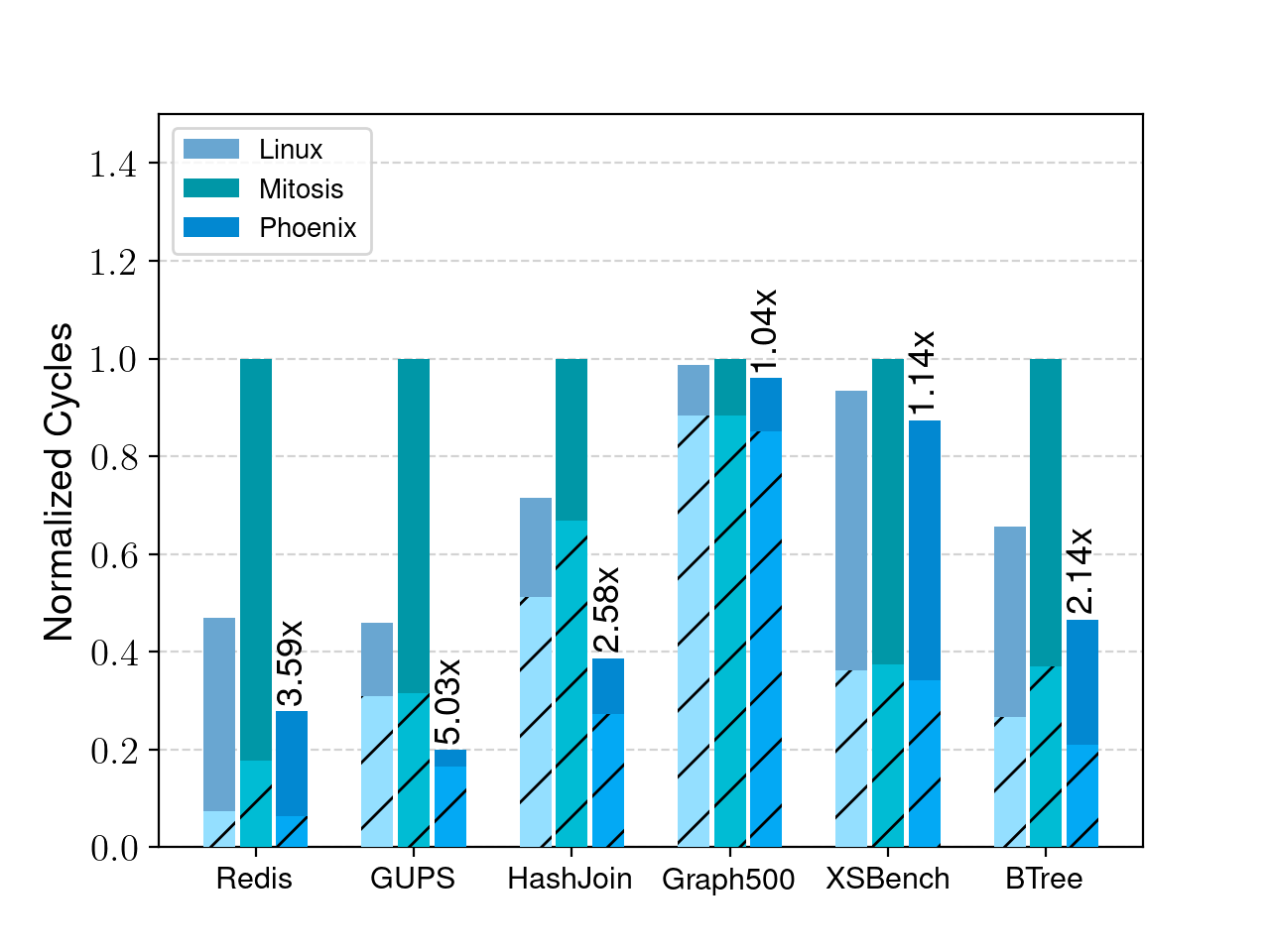}
        \caption{Memory bandwidth interference}
    \end{subfigure}
    \caption{Performance evaluation of Phoenix when a high-priority application shares its \emph{home node} with another application. Hashed part indicates page-walks cycles. The annotated enhancement is compared to Mitosis.}
    \label{fig:mba-benchmarks}
\end{figure*}

Co-locating applications on a node happens commonly in Phoenix due to threads consolidation. In such a scenario, contention may occur for memory bandwidth, decreasing the high-priority application performance. To resolve this, Phoenix uses Intel Memory Bandwidth Allocation (MBA) to throttle the memory bandwidth usage of the antagonistic application to the maximum possible. To evaluate this strategy, we configure benchmarks to spawn 16 threads while STREAM spawns 48. This configuration ensures that 16 threads of STREAM share the \emph{home node} of our targeted benchmarks utilizing all 64 cores of the system. Our analysis shows that 16 threads of STREAM are enough to saturate the memory bandwidth on our Skylake machine over $90\%$. As presented in Figure \ref{fig:mba-benchmarks}, we measured $1.95x$ improvement compared to Linux and $2.51x$ compared to Mitosis when UPI was saturated. Also, when memory bandwidth was throttled, we measured $1.55x$ and $2.58x$ performance improvement in Linux and Mitosis, correspondingly.

\subsection{Discussion}
\noindent\textbf{Linux implementation:} As a proof-of-concept, we implemented Phoenix as a loadable kernel module for Linux with minimal kernel changes on top of kernel v5.4 for the x86\_64 architecture. We chose Linux as a stable and widely used OS that enables Phoenix to be evaluated on top of a complex OS where much subtle interaction with components happens. 

\noindent\textbf{User-space delegation:} Implementing the Phoenix scheduling policy can be achieved by leveraging frameworks like Google ghOSt \cite{ghost}, which receive scheduling policies from user space. Similarly, the memory binding system calls available in Linux are sufficient for implementing the Phoenix memory management policy. Since the kernel is responsible for managing address space and page tables, the page table replication mechanism needs to be implemented within the kernel. However, in micro-kernels or multi-kernels (e.g., Barrelfish \cite{barrelfish}), where address space management is moved to the user space, the Phoenix replication mechanism can be linked as a LibraryOS without requiring any modifications to the kernel.

\noindent \textbf{Threshold finding:} Phoenix statically set a threshold determining the ratio of page-walk cycles to total CPU cycles on which it toggles Phoenix on and off for processes. Since we use \inlinecode{sysctl} to update this number, future works are enabled to a wide range of solutions, including but not limited to statistical or machine learning solutions.

\noindent\textbf{Limitation and future work:} In the context of socket consolidation, it is important to acknowledge that performance improvements are not guaranteed in all cases. One particular scenario that highlights this limitation is when an application's threads contend for shared resources. For instance, consider an application with \emph{N} threads that already saturate the memory bandwidth. If \emph{N+K} threads are consolidated on a single node, the increased thread count can exacerbate the contention for shared resources and potentially lead to degraded performance. This limitation emphasizes the need for further investigation and future work to dynamically resolve such resource contention issues.

Also, We believe other dynamic partitioning technologies, such as cache allocation technology (CAT), could be used to optimize Phoenix further for performance. We leave optimizations for future work.

\section{Related Work} \label{section:related-works}
In this section, we provide an overview of relevant works that have not been previously discussed. These works explore system models distinct from Phoenix and therefore were not included in our comparison.

Although several studies have suggested process placement strategies, none of them have specifically discussed optimizing scheduling by considering page table placement in NUMA systems. Our comparison focuses exclusively on Mitosis, as it is the only existing work that specifically addresses page table replication within the same system model as Phoenix.

\textbf{Tiered memory systems:} Tiered memory systems involve the migration of page tables between different memory tiers. Radiant is a cutting-edge solution for page table migration in tiered memory systems \cite{radiant}, specifically focusing on migration between DRAM and NVMM (Non-Volatile Memory Module). In Linux, when sufficient DRAM memory is available, it may allocate portions of the page table on slower NVMM. Radiant addresses this by selectively placing critical sections of the page table in fast DRAM, while migrating the remaining sections between NVMM and DRAM to achieve optimal memory placement. On average, Radiant reduces CPU cycles by $20\%$ compared to Linux.

\textbf{Virtualized systems:}
Virtualized systems utilize extended page tables that require two levels of address translations \cite{vmitosis}. The virtual address of the guest must be translated to the corresponding physical address. This physical address represents the virtual address of the host, which needs further translation to determine the host's physical address. Resolving an address through a 2D page-walk process requires $24$ memory accesses. vMitosis, similar to the Mitosis sibling, applies the same replication mechanism to virtual machines and has achieved a speedup of up to $1.6x$ compared to Linux. On the other hand, recent work \cite{translation-pass-through}, have explored a solution to use 1D address translation for virtual machines to achieve near-native translation latency. These findings could potentially complement the vMitosis approach.

\textbf{Flattening the page table:}
A promising strategy to alleviate page-walk overhead involves minimizing memory accesses, as explored in recent studies \cite{every-walk-is-a-hit, compendia}. These works propose transforming 4 KB page table pages into 2 MB huge pages, effectively decreasing the required levels of page tables from four to two and reducing memory access requirements to nearly one in many scenarios. However, since the hardware unit responsible for performing page-walks is designed based on current structures, implementing such flattening approaches necessitates corresponding changes in hardware.

\textbf{Kernel state replication:} NrOS, a state-of-the-art kernel state replication system, offers a unique approach to addressing scalability challenges in operating system kernels \cite{nros}. Developed from scratch using the Rust programming language, NrOS consists of approximately $11,000$ lines of kernel code and an additional $16,000$ lines of code for kernel libraries, including drivers and the bootloader. While NrOS shares similarities with Mitosis in replicating page tables across NUMA nodes, it distinguishes itself through the adoption of a novel approach called \emph{node replication}. This technique ensures coherency maintenance and enhances scalability. However, it should be noted that NrOS's current destructive implementation is incompatible with Linux and legacy applications, limiting its practicality in present-day data centers. Alternatively, Phoenix emerges as a promising production solution due to its compatibility with legacy applications, minimal kernel modifications, and reliance on stable kernel building blocks.

\section{Conclusion} \label{section:conclusion}
This paper presents Phoenix, a novel on-demand page table replication with a new process placement policy that achieves better performance and QoS where multiple applications are co-located on NUMA systems. Phoenix maximizes the locality of page table and process via replication and consolidation techniques while removing memory bandwidth contention to achieve the fastest page-walk. An evaluation of a range of real-world applications shows that Phoenix achieves its goals in its targeted environment. Moreover, we made a case for taking both process and page table placement into a primary OS design consideration with solid evidence. We plan to open-source the kernel and LKM used in this study upon cleaning the code to inspire future research on schedulers and memory managers that are aware of each other.

Each workload performs differently under various CPU scheduling and memory management policies. To fulfill this requirement, recent works proposed solutions to apply user-defined CPU scheduling policies in data centers \cite{ghost}. We believe our memory management and CPU co-scheduling design could be pushed further as a research line by considering a broader range of workloads and applying dynamic user-defined memory management and CPU co-scheduling policies as a trending research in the field.




\bibliographystyle{unsrt}
\bibliography{sample-base}

\begin{thebibliography}{10}

\bibitem{asymmetry-matters}
Baptiste Lepers, Vivien Quema, and Alexandra Fedorova.
\newblock Thread and memory placement on {NUMA} systems: Asymmetry matters.
\newblock In {\em 2015 USENIX Annual Technical Conference (USENIX ATC 15)}, pages 277--289, Santa Clara, CA, July 2015. USENIX Association.

\bibitem{page-management-design-space}
Jonghyeon Kim, Wonkyo Choe, and Jeongseob Ahn.
\newblock Exploring the design space of page management for {Multi-Tiered} memory systems.
\newblock In {\em 2021 USENIX Annual Technical Conference (USENIX ATC 21)}, pages 715--728. USENIX Association, July 2021.

\bibitem{balm}
D.~Gureya, Vladimir Vlassov, and J.~Barreto.
\newblock Brief announcement : Balm: qos-aware memory bandwidth partitioning for multi-socket cloud nodes.
\newblock In {\em Annual ACM Symposium on Parallelism in Algorithms and Architectures :}, pages 435--438. Association for Computing Machinery (ACM), 2021.
\newblock Part of proceedings ISBN: 978-1-4503-8070-6QC 20220330.

\bibitem{mitosis}
Reto Achermann, Ashish Panwar, Abhishek Bhattacharjee, Timothy Roscoe, and Jayneel Gandhi.
\newblock Mitosis: Transparently self-replicating page-tables for large-memory machines.
\newblock In {\em Proceedings of the Twenty-Fifth International Conference on Architectural Support for Programming Languages and Operating Systems}, ASPLOS '20, page 283–300, New York, NY, USA, 2020. Association for Computing Machinery.

\bibitem{vmitosis}
Ashish Panwar, Reto Achermann, Arkaprava Basu, Abhishek Bhattacharjee, K.~Gopinath, and Jayneel Gandhi.
\newblock Fast local page-tables for virtualized numa servers with vmitosis.
\newblock In {\em Proceedings of the 26th ACM International Conference on Architectural Support for Programming Languages and Operating Systems}, ASPLOS '21, page 194–210, New York, NY, USA, 2021. Association for Computing Machinery.

\bibitem{radiant}
Sandeep Kumar, Aravinda Prasad, Smruti~R. Sarangi, and Sreenivas Subramoney.
\newblock Radiant: Efficient page table management for tiered memory systems.
\newblock In {\em Proceedings of the 2021 ACM SIGPLAN International Symposium on Memory Management}, ISMM 2021, page 66–79, New York, NY, USA, 2021. Association for Computing Machinery.

\bibitem{nros}
Ankit Bhardwaj, Chinmay Kulkarni, Reto Achermann, Irina Calciu, Sanidhya Kashyap, Ryan Stutsman, Amy Tai, and Gerd Zellweger.
\newblock Nros: Effective replication and sharing in an operating system.
\newblock In {\em 15th {USENIX} Symposium on Operating Systems Design and Implementation ({OSDI} 21)}, pages 295--312. {USENIX} Association, July 2021.

\bibitem{nuKSM}
Akash Panda, Ashish Panwar, and Arkaprava Basu.
\newblock nuksm: Numa-aware memory de-duplication on multi-socket servers.
\newblock In Jaejin Lee and Albert Cohen, editors, {\em 30th International Conference on Parallel Architectures and Compilation Techniques, {PACT} 2021, Atlanta, GA, USA, September 26-29, 2021}, pages 258--273. {IEEE}, 2021.

\bibitem{skip-don't-walk}
Thomas~W. Barr, Alan~L. Cox, and Scott Rixner.
\newblock Translation caching: Skip, don't walk (the page table).
\newblock In {\em Proceedings of the 37th Annual International Symposium on Computer Architecture}, ISCA '10, page 48–59, New York, NY, USA, 2010. Association for Computing Machinery.

\bibitem{is-data-migration-evil-on-nvm}
Jungwook Han, Hongsu Byun, Hyungjoon Kwon, Sungyong Park, and Youngjae Kim.
\newblock Is data migration evil in the nvm file system?
\newblock In {\em 2021 IEEE International Conference on Autonomic Computing and Self-Organizing Systems Companion (ACSOS-C)}, pages 26--31, 2021.

\bibitem{inter-node-load-balancing-for-numa-systems}
Mei-Ling Chiang and Wei-Lun Su.
\newblock Thread-aware mechanism to enhance inter-node load balancing for multithreaded applications on numa systems.
\newblock {\em Applied Sciences}, 11(14), 2021.

\bibitem{locality-vs-balance}
Matthias Diener, Eduardo~H.M. Cruz, and Philippe~O.A. Navaux.
\newblock Locality vs. balance: Exploring data mapping policies on numa systems.
\newblock In {\em 2015 23rd Euromicro International Conference on Parallel, Distributed, and Network-Based Processing}, pages 9--16, 2015.

\bibitem{nimble}
Zi~Yan, Daniel Lustig, David Nellans, and Abhishek Bhattacharjee.
\newblock Nimble page management for tiered memory systems.
\newblock In {\em Proceedings of the Twenty-Fourth International Conference on Architectural Support for Programming Languages and Operating Systems}, ASPLOS '19, page 331–345, New York, NY, USA, 2019. Association for Computing Machinery.

\bibitem{numa-aware-data-placement-in-hybrid-memories}
Zhuohui Duan, Haikun Liu, Xiaofei Liao, Hai Jin, Wenbin Jiang, and Yu~Zhang.
\newblock Hinuma: Numa-aware data placement and migration in hybrid memory systems.
\newblock In {\em 2019 IEEE 37th International Conference on Computer Design (ICCD)}, pages 367--375, 2019.

\bibitem{numa-aware-core-allocation}
Jiri Dokulil and Siegfried Benkner.
\newblock Numa-aware cpu core allocation in cooperating dynamic applications.
\newblock In {\em 2020 IEEE International Parallel and Distributed Processing Symposium Workshops (IPDPSW)}, pages 950--957, 2020.

\bibitem{traffic-management}
Mohammad Dashti, Alexandra Fedorova, Justin Funston, Fabien Gaud, Renaud Lachaize, Baptiste Lepers, Vivien Quema, and Mark Roth.
\newblock Traffic management: A holistic approach to memory placement on numa systems.
\newblock {\em SIGARCH Comput. Archit. News}, 41(1):381–394, mar 2013.

\bibitem{challenges-of-modern-numa-scheduling}
Fabien Gaud, Baptiste Lepers, Justin Funston, Mohammad Dashti, Alexandra Fedorova, Vivien Qu\'{e}ma, Renaud Lachaize, and Mark Roth.
\newblock Challenges of memory management on modern numa systems.
\newblock {\em Commun. ACM}, 58(12):59–66, nov 2015.

\bibitem{trident-tiered-storage}
Herodotos Herodotou and Elena Kakoulli.
\newblock Trident: Task scheduling over tiered storage systems in big data platforms.
\newblock {\em Proc. VLDB Endow.}, 14(9):1570–1582, may 2021.

\bibitem{task-based-runtime-systems}
Marcos Maroñas, Antoni Navarro, Eduard Ayguadé, and Vicenç Beltran.
\newblock Mitigating the numa effect on task-based runtime systems.
\newblock {\em The Journal of Supercomputing}, pages 1--26, 04 2023.

\bibitem{vtmm}
Sai Sha, Chuandong Li, Yingwei Luo, Xiaolin Wang, and Zhenlin Wang.
\newblock Vtmm: Tiered memory management for virtual machines.
\newblock In {\em Proceedings of the Eighteenth European Conference on Computer Systems}, EuroSys '23, page 283–297, New York, NY, USA, 2023. Association for Computing Machinery.

\bibitem{memory-disaggregation}
Bin Gao, Hao-Wei Tee, Alireza Sanaee, Soh~Boon Jun, and Djordje Jevdjic.
\newblock Os-level implications of using dram caches in memory disaggregation.
\newblock In {\em 2022 IEEE International Symposium on Performance Analysis of Systems and Software (ISPASS)}, pages 153--155, 2022.

\bibitem{hydra}
Bin Gao, Qingxuan Kang, Hao-Wei Tee, Kyle Timothy~Ng Chu, Alireza Sanaee, and Djordje Jevdjic.
\newblock Scalable and effective page-table and {TLB} management on {NUMA} systems.
\newblock In {\em 2024 USENIX Annual Technical Conference (USENIX ATC 24)}, pages 445--461, Santa Clara, CA, July 2024. USENIX Association.

\bibitem{dancing-in-the-dark}
Jinyoung Choi, Sergey Blagodurov, and Hung-Wei Tseng.
\newblock Dancing in the dark: Profiling for tiered memory.
\newblock In {\em 2021 IEEE International Parallel and Distributed Processing Symposium (IPDPS)}, pages 13--22, 2021.

\bibitem{esxi-scheduler}
Xunjia Lu.
\newblock Extreme performance series: vsphere compute \& memory schedulers.
\newblock \url{https://magander.se/vmworld-2017-breakout-session-extreme-performance-series-vsphere-compute-memory-schedulers/}, 2017.

\bibitem{caladan}
Joshua Fried, Zhenyuan Ruan, Amy Ousterhout, and Adam Belay.
\newblock Caladan: Mitigating interference at microsecond timescales.
\newblock In {\em 14th USENIX Symposium on Operating Systems Design and Implementation (OSDI 20)}, pages 281--297. USENIX Association, November 2020.

\bibitem{emba}
Yaocheng Xiang, Chencheng Ye, Xiaolin Wang, Yingwei Luo, and Zhenlin Wang.
\newblock Emba: Efficient memory bandwidth allocation to improve performance on intel commodity processor.
\newblock In {\em Proceedings of the 48th International Conference on Parallel Processing}, ICPP 2019, New York, NY, USA, 2019. Association for Computing Machinery.

\bibitem{micro-architecture-interference}
Haecheon Kim, Seungmin Lim, Junkee Yoon, Seungjae Baek, Jongmoo Choi, and Seong-je Cho.
\newblock Analysis of micro-architecture resources interference on multicore numa systems.
\newblock In {\em Proceedings of the 31st Annual ACM Symposium on Applied Computing}, SAC '16, page 1871–1876, New York, NY, USA, 2016. Association for Computing Machinery.

\bibitem{shenango}
Amy Ousterhout, Joshua Fried, Jonathan Behrens, Adam Belay, and Hari Balakrishnan.
\newblock Shenango: Achieving high {CPU} efficiency for latency-sensitive datacenter workloads.
\newblock In {\em 16th USENIX Symposium on Networked Systems Design and Implementation (NSDI 19)}, pages 361--378, Boston, MA, February 2019. USENIX Association.

\bibitem{lego-os}
Yizhou Shan, Yutong Huang, Yilun Chen, and Yiying Zhang.
\newblock {LegoOS}: A disseminated, distributed {OS} for hardware resource disaggregation.
\newblock In {\em 13th USENIX Symposium on Operating Systems Design and Implementation (OSDI 18)}, pages 69--87, Carlsbad, CA, October 2018. USENIX Association.

\bibitem{stream}
John~D. McCalpin.
\newblock Stream: Sustainable memory bandwidth in high performance computers.
\newblock \url{https://www.cs.virginia.edu/stream/}, 2022.

\bibitem{apache}
Apache~Software Foundation.
\newblock Apache http server project.
\newblock \url{https://httpd.apache.org/}, 2022.

\bibitem{mapreduce}
Jeffrey Dean and Sanjay Ghemawat.
\newblock Mapreduce: Simplified data processing on large clusters.
\newblock {\em Commun. ACM}, 51(1):107–113, jan 2008.

\bibitem{evaluating-mapreduce}
Colby Ranger, Ramanan Raghuraman, Arun Penmetsa, Gary Bradski, and Christos Kozyrakis.
\newblock Evaluating mapreduce for multi-core and multiprocessor systems.
\newblock In {\em 2007 IEEE 13th International Symposium on High Performance Computer Architecture}, pages 13--24, 2007.

\bibitem{didi}
Carlos Villavieja, Vasileios Karakostas, Lluis Vilanova, Yoav Etsion, Alex Ramirez, Avi Mendelson, Nacho Navarro, Adrian Cristal, and Osman~S. Unsal.
\newblock Didi: Mitigating the performance impact of tlb shootdowns using a shared tlb directory.
\newblock In {\em 2011 International Conference on Parallel Architectures and Compilation Techniques}, pages 340--349, 2011.

\bibitem{optimizing-tlb-shootdown-algorithm}
Nadav Amit.
\newblock Optimizing the {TLB} shootdown algorithm with page access tracking.
\newblock In {\em 2017 USENIX Annual Technical Conference (USENIX ATC 17)}, pages 27--39, Santa Clara, CA, July 2017. USENIX Association.

\bibitem{latr}
Mohan~Kumar Kumar, Steffen Maass, Sanidhya Kashyap, J\'{a}n Vesel\'{y}, Zi~Yan, Taesoo Kim, Abhishek Bhattacharjee, and Tushar Krishna.
\newblock Latr: Lazy translation coherence.
\newblock In {\em Proceedings of the Twenty-Third International Conference on Architectural Support for Programming Languages and Operating Systems}, ASPLOS '18, page 651–664, New York, NY, USA, 2018. Association for Computing Machinery.

\bibitem{analysis-of-linux-scalability-on-many-cores}
Silas Boyd-Wickizer, Austin~T. Clements, Yandong Mao, Aleksey Pesterev, M.~Frans Kaashoek, Robert Morris, and Nickolai Zeldovich.
\newblock An analysis of linux scalability to many cores.
\newblock In {\em Proceedings of the 9th USENIX Conference on Operating Systems Design and Implementation}, OSDI'10, page 1–16, USA, 2010. USENIX Association.

\bibitem{radixvm}
Austin~T. Clements, M.~Frans Kaashoek, and Nickolai Zeldovich.
\newblock Radixvm: Scalable address spaces for multithreaded applications.
\newblock In {\em Proceedings of the 8th ACM European Conference on Computer Systems}, EuroSys '13, page 211–224, New York, NY, USA, 2013. Association for Computing Machinery.

\bibitem{optimizing-mapreduce}
Yandong Mao, Robert Morris, and Frans Kaashoek.
\newblock Optimizing mapreduce for multicore architectures.
\newblock Technical report, Technical Report MIT-CSAIL-TR-2010-020, MIT, 2010.

\bibitem{every-walk-is-a-hit}
Chang~Hyun Park, Ilias Vougioukas, Andreas Sandberg, and David Black-Schaffer.
\newblock Every walk’s a hit: Making page walks single-access cache hits.
\newblock In {\em Proceedings of the 27th ACM International Conference on Architectural Support for Programming Languages and Operating Systems}, ASPLOS '22, page 128–141, New York, NY, USA, 2022. Association for Computing Machinery.

\bibitem{compendia}
Sam Ainsworth and Timothy~M. Jones.
\newblock Compendia: Reducing virtual-memory costs via selective densification.
\newblock In {\em Proceedings of the 2021 ACM SIGPLAN International Symposium on Memory Management}, ISMM 2021, page 52–65, New York, NY, USA, 2021. Association for Computing Machinery.

\bibitem{big-memory-servers}
Arkaprava Basu, Jayneel Gandhi, Jichuan Chang, Mark~D. Hill, and Michael~M. Swift.
\newblock Efficient virtual memory for big memory servers.
\newblock In {\em Proceedings of the 40th Annual International Symposium on Computer Architecture}, ISCA '13, page 237–248, New York, NY, USA, 2013. Association for Computing Machinery.

\bibitem{trident-harnessing-architectural-resources}
Venkat Sri~Sai Ram, Ashish Panwar, and Arkaprava Basu.
\newblock Trident: Harnessing architectural resources for all page sizes in x86 processors.
\newblock In {\em MICRO-54: 54th Annual IEEE/ACM International Symposium on Microarchitecture}, MICRO '21, page 1106–1120, New York, NY, USA, 2021. Association for Computing Machinery.

\bibitem{adaptive-page-migration}
Taekyung Heo, Yang Wang, Wei Cui, Jaehyuk Huh, and Lintao Zhang.
\newblock Adaptive page migration policy with huge pages in tiered memory systems.
\newblock {\em IEEE Transactions on Computers}, 71(1):53--68, 2022.

\bibitem{decade-of-wasted-cores}
Jean-Pierre Lozi, Baptiste Lepers, Justin Funston, Fabien Gaud, Vivien Qu\'{e}ma, and Alexandra Fedorova.
\newblock The linux scheduler: A decade of wasted cores.
\newblock In {\em Proceedings of the Eleventh European Conference on Computer Systems}, EuroSys '16, New York, NY, USA, 2016. Association for Computing Machinery.

\bibitem{battle-of-schedulers}
Justinien Bouron, Sebastien Chevalley, Baptiste Lepers, Willy Zwaenepoel, Redha Gouicem, Julia Lawall, Gilles Muller, and Julien Sopena.
\newblock The battle of the schedulers: {FreeBSD} {ULE} vs. linux {CFS}.
\newblock In {\em 2018 USENIX Annual Technical Conference (USENIX ATC 18)}, pages 85--96, Boston, MA, July 2018. USENIX Association.

\bibitem{ptlbmalloc2}
Stijn Schildermans, Kris Aerts, Jianchen Shan, and Xiaoning Ding.
\newblock Ptlbmalloc2: Reducing tlb shootdowns with high memory efficiency.
\newblock {\em 2020 IEEE Intl Conf on Parallel \& Distributed Processing with Applications, Big Data \& Cloud Computing, Sustainable Computing \& Communications, Social Computing \& Networking (ISPA/BDCloud/SocialCom/SustainCom)}, pages 76--83, 2020.

\bibitem{intel-rdt}
Intel.
\newblock Resource director technology.
\newblock \url{https://www.intel.com/content/www/us/en/architecture-and-technology/resource-director-technology.html}, 2023.

\bibitem{intel-mba}
Intel.
\newblock Memory bandwidth allocation.
\newblock \url{https://www.intel.com/content/www/us/en/developer/articles/technical/introduction-to-memory-bandwidth-allocation.html}, 2023.

\bibitem{holmes}
Aidi Pi, Xiaobo Zhou, and Chengzhong Xu.
\newblock Holmes: Smt interference diagnosis and cpu scheduling for job co-location.
\newblock In {\em Proceedings of the 31st International Symposium on High-Performance Parallel and Distributed Computing}, HPDC '22, page 110–121, New York, NY, USA, 2022. Association for Computing Machinery.

\bibitem{pmctrack}
J.~C. Saez, A.~Pousa, R.~Rodriíguez-Rodriíguez, F.~Castro, and M.~Prieto-Matias.
\newblock {PMCTrack: Delivering Performance Monitoring Counter Support to the OS Scheduler}.
\newblock {\em The Computer Journal}, 60(1):60--85, 01 2017.

\bibitem{redis}
Redis Labs.
\newblock Redis.
\newblock \url{https://redis.io}, 2023.

\bibitem{gups}
HPCCHALLENGE.
\newblock Randomaccess: Gups (giga updates per second).
\newblock \url{https://icl.utk.edu/projectsfiles/hpcc/RandomAccess/.}, 2023.

\bibitem{xsbench}
John~R Tramm, Andrew~R Siegel, Tanzima Islam, and Martin Schulz.
\newblock Xsbench-the development and verification of a performance abstraction for monte carlo reactor analysis.
\newblock {\em The Role of Reactor Physics toward a Sustainable Future (PHYSOR)}, 2014.

\bibitem{graph500}
James~A. Ang, Brian~W. Barrett, Kyle~B. Wheeler, and Richard~C. Murphy.
\newblock Introducing the graph 500.
\newblock 2010.

\bibitem{ghost}
Jack~Tigar Humphries, Neel Natu, Ashwin Chaugule, Ofir Weisse, Barret Rhoden, Josh Don, Luigi Rizzo, Oleg Rombakh, Paul Turner, and Christos Kozyrakis.
\newblock ghost: Fast \& flexible user-space delegation of linux scheduling.
\newblock In {\em Proceedings of the ACM SIGOPS 28th Symposium on Operating Systems Principles}, pages 588--604, 2021.

\bibitem{barrelfish}
Andrew Baumann, Paul Barham, Pierre-Evariste Dagand, Tim Harris, Rebecca Isaacs, Simon Peter, Timothy Roscoe, Adrian Sch{\"u}pbach, and Akhilesh Singhania.
\newblock The multikernel: a new os architecture for scalable multicore systems.
\newblock In {\em Proceedings of the ACM SIGOPS 22nd symposium on Operating systems principles}, pages 29--44, 2009.

\bibitem{translation-pass-through}
Shai Bergman, Mark Silberstein, Takahiro Shinagawa, Peter Pietzuch, and Lluís Vilanova.
\newblock Translation pass-through for near-native paging performance in vms.
\newblock In {\em 2023 USENIX Annual Technical Conference (USENIX ATC 23)}, BOSTON, MA, USA, July 2023. USENIX Association.

\end{thebibliography}

\end{document}